# Presolar O- and C-anomalous grains in unequilibrated ordinary chondrite matrices

Jens Barosch[1*], Larry R. Nittler[1], Jianhua Wang[1], Elena Dobrică[2], Adrian J. Brearley[3], Dominik C. Hezel[4], Conel M. O'D. Alexander[1]

[1]Earth and Planets Laboratory, Carnegie Institution of Washington, 5241 Broad Branch Rd. NW, Washington DC 20015, USA.

[2]Hawai'i Institute of Geophysics and Planetology, University of Hawai'i at Mānoa, 1680 East-West Road, Honolulu 96822, USA.

[3]Department of Earth and Planetary Sciences, 1 University of New Mexico, Albuquerque, New Mexico 87131-0001, USA.

[4]Institut für Geowissenschaften, Goethe-Universität Frankfurt, Altenhöferallee 1, 60438 Frankfurt am Main, Germany.

**corresponding author:* jbarosch@carnegiescience.edu

Methods – Results – Discussion – Conclusions – Tables – Figures

**Abstract**

Presolar grains are trace components in chondrite matrices. Their abundances and compositions have been systematically studied in carbonaceous chondrites but rarely in situ in other major chondrite classes. We have conducted a NanoSIMS isotopic search for presolar grains with O- and C-anomalous isotopic compositions in the matrices of the unequilibrated ordinary chondrites Semarkona (LL3.00), Meteorite Hills 00526 (L/LL3.05), and Northwest Africa 8276 (L3.00). The matrices of even the most primitive ordinary chondrites have been aqueously altered and/or thermally metamorphosed, destroying their presolar grain populations to varying extents. In addition to randomly placed isotope maps, we specifically targeted recently reported, relatively pristine Semarkona matrix areas to better explore the original inventory of presolar grains in this meteorite. In all samples, we found a total of 122 O-anomalous grains (silicates + oxides), 79 SiC grains, and 22 C-anomalous carbonaceous grains (organics, graphites). Average matrix-normalized abundances with 1σ uncertainties are $151\,^{+50}_{-46}$ ppm O-anomalous grains, $53\,^{+14}_{-12}$ ppm SiC grains and $56\,^{+19}_{-14}$ ppm carbonaceous grains in Semarkona, $55\,^{+11}_{-10}$ ppm (O-anom.), $22\,^{+5}_{-4}$ ppm (SiC) and $3\,^{+2}_{-1}$ ppm (carb.) in MET 00526 and $12\,^{+6}_{-3}$ ppm (O-anom.), $15\,^{+7}_{-5}$ ppm (SiC) and $1\,^{+3}_{-1}$ ppm (carb.) in NWA 8276. In relatively pristine ordinary chondrites and in primitive carbonaceous and C-ungrouped chondrites, the O and C isotopic composition of presolar grains and their matrix-normalized abundances are similar, despite the likely differences in chondrite-formation time and nebular location. These results suggest a relatively homogenous distribution of presolar dust across major chondrite-forming reservoirs in the solar nebula. Secondary asteroidal processes are mainly responsible for differences in presolar grain abundances between and within chondrites, highlighting the need to identify and target the most pristine chondrite matrices for such studies.

## 1. Introduction

Presolar stardust grains are nm- to µm-sized particles that formed around dying stars before the birth of the Solar System. They occur in the fine-grained matrix of primitive chondrites as trace components and are identified by their highly anomalous isotopic compositions, which largely reflect nucleosynthesis in their parent stars. Their study allows insights into the conditions and nucleosynthetic processes in their parent stars and helps to constrain the evolution of the early Solar System (cf. Zinner 2014, and references therein). High-resolution in situ nanoscale secondary ion mass spectrometry (NanoSIMS) studies of the abundances and characteristics of presolar grains still contained within their host meteorites are particularly useful for determining the properties of presolar materials in the solar nebula. Furthermore, such studies are crucial for identifying possible heterogeneities in the distribution of presolar materials in the protoplanetary disk and trace secondary asteroidal processes that altered chondrite matrices.

The known presolar grains are primarily either O-rich or C-rich phases. O-rich presolar grains are generally identified by their anomalous O-isotopic compositions compared to materials that formed in the protoplanetary disk. Most grains are silicates (e.g., olivines and pyroxenes, but also non-stoichiometric and/or non-crystalline phases; Floss and Haenecour 2016a) or oxides (e.g., hibonite, spinel, corundum, magnetite, rutile, chromite; Zinner 2014). O-anomalous grains have been subdivided into four groups in the O three-isotope plot, linking them to their likely stellar origins (Nittler et al. 1997, 2008; Floss and Haenecour 2016a): Group 1 grains are $^{17}$O-rich and most likely formed in low-mass asymptotic giant branch (AGB) stars with close-to-solar metallicities, although supernovae have been recently suggested as sources for a $^{25}$Mg-enriched sub-set of them (Leitner and Hoppe 2019; Hoppe et al. 2021). Group 2 grains also have elevated $^{17}$O/$^{16}$O but are highly depleted in $^{18}$O, whereas Group 3 grains are depleted in $^{17}$O and $^{18}$O. Both groups are also associated with AGB stars and either experienced non-standard mixing processes (Group 2) or formed in stars with subsolar metallicities (Group 3). Group 4 grains are enriched in $^{18}$O and may have formed in supernovae. In addition, a small number of ungrouped grains with extreme $^{17}$O or $^{16}$O enrichments likely formed in novae and supernovae, respectively (Nittler et al. 1998; Gyngard et al. 2010).

The best-studied C-rich presolar phase is silicon carbide (SiC). Depending on their C-, N- and Si-isotopic compositions, SiC grains have been classified as mainstream, AB, C, N, X, Y, or Z grains. Mainstream grains are the most common (>90%; Zinner 2014, and references therein). They have $^{12}$C/$^{13}$C ratios that are typically between 12–100, correlated $\delta^{29}$Si and $\delta^{30}$Si values with a slope of 1.37 (Lugaro et al. 1999; Stephan et al. 2020), and typically $^{14}N$ enrichments relative to solar. The origin of mainstream, Y, and Z grains is linked to AGB stars (Zinner 2014, and references therein). The other SiC groups are the ejecta of novae (N), supernovae (X, C), or have still ambiguous origins (AB). New multi-element isotopic data, as well as cluster analysis of existing data will help to refine this classification by better linking SiC grains to their parent stars (Liu et al. 2019, 2021; Boujibar et al. 2021).

An additional type of isotopically anomalous, possibly presolar, material found in meteorites is macromolecular organic matter, the dominant form of C in chondrites (Alexander et al. 2017a). Large $^{2}H$ and $^{15}N$ excesses in this material point to very low-T chemistry and a possible origin in the presolar molecular cloud or the cold, radiation-rich outer Solar System. While most of the meteoritic organic matter has $^{12}C/^{13}C$ ratios within the range of Solar System materials, some organic grains with $^{13}C$ enrichments and/or depletions are present, though the anomalies are less extreme than those found in most presolar stardust (Floss and Stadermann 2009a; Alexander et al. 2017a; Nittler et al. 2018a).

The matrix-normalized abundances of presolar O-anomalous grains in the most primitive carbonaceous chondrites (CCs) are typically in the range of the higher tens to the lower hundreds of parts per million (ppm). The highest abundances observed to date are ~240 ppm in Dominion Range (DOM) 08006 (CO3.0; Haenecour et al. 2018; Nittler et al. 2018a) and Asuka 12169 (CM3.0; Nittler et al. 2021). Presolar SiC grains in CC are typically in the range of 10–60 ppm (Davidson et al. 2014; Nittler et al. 2021). Presolar O-anomalous and SiC grains are also present in unequilibrated ordinary chondrites (UOC; e.g., Huss 1990; Huss and Lewis 1995; Alexander et al. 1990; Alexander 1993; Nittler et al. 1994, 1997; Mostefaoui et al. 2003, 2004; Davidson et al. 2014; Hoppe et al. 2021). However, a systematic in situ study of their abundances and compositions is missing. The few abstracts available reported very diverse in situ O-anomalous presolar grain abundances, ranging from 4–15 ppm in Semarkona (Mostefaoui et al. 2003; Tonotani et al. 2006) to up to 275 ppm in Meteorite Hills (MET) 00526 (Floss and Haenecour 2016b). SiC abundances are in the range of <1 ppm to 125 ppm (Tonotani et al. 2006; Floss and Haenecour 2016c).

A major challenge when determining the abundances of presolar grains is that only the most pristine matrices may preserve them in their original abundances. Presolar SiC and $Al_2O_3$ are susceptible to thermal metamorphism but relatively resilient to aqueous alteration, whereas presolar silicates are easily destroyed by all asteroidal processes (Floss and Haenecour 2016a). Choosing samples of low petrologic types is crucial. However, even these may not be sufficiently pristine: Semarkona is classified as LL3.0, but its matrix has been extensively aqueously altered (Alexander et al. 1989), explaining the almost complete absence of presolar (silicate) grains in the studies of Mostefaoui et al. (2003) and Tonotani et al. (2006). A recent transmission electron microscopy (TEM) study by Dobrică and Brearley (2020a) found that the extent of aqueous alteration of Semarkona matrix can be quite heterogeneous. They identified amorphous silicate-rich matrix domains, indicating that in these areas minimal alteration has occurred (Fig. 1). Relatively pristine, amorphous silicate-rich matrix areas are prime targets to examine as best as possible the pre-alteration inventory of presolar grains in UOCs. A similar TEM survey was conducted on MET 00526 (Dobrică and Brearley 2020b). The authors detected amorphous silicates but also veins of FeO-rich olivines and coarse-grained phyllosilicates, indicating that these regions might have experienced a higher degree of aqueous alteration than the more pristine parts of Semarkona's matrix. Additional challenges of studying presolar grains in UOCs arise from the low abundances of matrix (UOCs: typically <20 vol.%, CCs: variable but often >50 vol.%; Scott and Krot, 2014, and references therein) and the often clast-rich nature of matrix material.

Raster imaging with the NanoSIMS is the method of choice to identify and quantify presolar grains in situ. However, there are limitations to the method. Particularly problematic are the effects of long tails in the roughly Gaussian-shaped primary ion beam, which lead to contributions of neighboring materials to the signal in any given pixel. This "isotopic dilution" may prevent accurate determination of presolar grain sizes and/or isotopic compositions (e.g., Nguyen et al. 2007). Image simulations have been used previously to explore, and in some cases correct for, isotopic dilution in NanoSIMS images (Nguyen et al. 2007; Qin et al. 2011; Hoppe et al. 2018). Here, we expand on this work with new simulations (Supplementary Material). These will help to better understand the combined effects of beam tailing and counting statistics on the identification and quantification of presolar grains.

We present the first comprehensive and systematic in situ study of the abundances and compositions of presolar O- and C-anomalous grains in primitive UOCs. The samples investigated are Semarkona (LL3.00), MET 00526 (L/LL3.05), and Northwest Africa (NWA) 8276 (L3.00). In addition to randomly placed NanoSIMS maps, we specifically targeted amorphous silicate-rich domains in the matrix of Semarkona that only showed minimal evidence for aqueous alteration (Dobrică and Brearley 2020a, b). Thus, we were able to better estimate the original abundances of presolar grains in this sample. Particularly large, interesting, or unusual grains from this study were further investigated by TEM in companion studies by Singerling et al. (2022). We use our dataset to compare the presolar grain populations of UOCs and CCs and identify similarities and differences.

## 2. Methods

We used the CAMECA NanoSIMS 50L ion microprobe at the Carnegie Institution for in situ identification of O- and C-anomalous grains. Ion images of $^{12}C^-$, $^{13}C^-$, $^{16}O^-$, $^{17}O^-$, $^{18}O^-$, $^{28}Si^-$, $^{27}Al^{16}O^-$, and secondary electrons were recorded in multicollection mode by rastering the $Cs^+$ primary ion beam (~100–120 nm, ~0.5 pA) over contiguous 10×10 µm-sized areas. The mass resolving power on individual electron multipliers (EM) was high enough to separate isobaric interferences (e.g., $^{16}OH$ on $^{17}O$). The peak positions of $^{12}C^-$ and $^{16}O^-$ were automatically monitored every five cycles and, additionally, a Nuclear Magnetic Resonance probe was used to maintain peak stability. Each area was analyzed for 25–30 sequential cycles. The resolution of each image frame was 256×256 pixels with a 1500 µs counting time per pixel per cycle.

We used the L'image software for data reduction as described by Nittler et al. (2018a). All images were autocorrected for a 44 ns EM deadtime and for shifts between individual ion image frames. Isotopic images of $^{16}O$ were also empirically corrected for the QSA effect (Slodzian et al. 2004; Ogliore et al. 2021) based on linear fits to the average pixel $^{18}O/^{16}O$ ratio as a function of $^{16}O$ count rate. QSA corrections were on average <5%. We normalized the O and C isotopic ratios of each map to that of the meteorite's matrix. The O- and C-anomalous grain candidates were then identified in sigma images of $\delta^{13}C$, $\delta^{17}O$ and $\delta^{18}O$, where every pixel represents the number of standard deviations from the matrix values (see footnote A for δ-value calculation). The errors were determined from Poisson statistics, i.e., the standard deviation is derived from the square root of the total number of counted secondary ions except that for isotopic depletions it is calculated from the total counts expected for a terrestrial isotopic ratio (cf. Nittler et al. 2018a).

Selected regions of interests (ROIs = grain candidates) consist of several contiguous pixels with isotopic ratios that are statistically significantly different (see below) from the average ratios of the matrix in the ion images. As in recent publications (Nittler et al. 2018a, 2019, 2021), we generally defined ROIs by including all pixels within the full width at half maximum of the anomalous region(s). Because the most central pixels of ROIs are more anomalous than pixels at the ROI borders, due to primary ion beam tailing, our ROIs only provide a lower limit to the anomalies. However, due to isotopic dilution (Supplementary Material), this is also true if only the central pixels are considered for most grains and our approach has the advantage of providing a self-consistent and uniform method for all grains.

We note that most published in situ presolar grain studies have not described in detail the procedure(s) used to select which pixels are included in a presolar grain ROI.

The ROIs were considered to be O-anomalous grains if their isotopic compositions were outside the range of Solar System materials by at least 5σ (e.g., Fig. 1a) and they had a minimum size of nine pixels (≈130 nm in diameter). The 5σ criterion is based on NanoSIMS image simulations which indicate that false candidates arise below this significance level due to counting statistics (Supplementary Material). It is more stringent than used in our previous work and is similar to the 5.3σ threshold advocated by Hoppe et al. (2015) based on measurements of a terrestrial analog material. We attempted to determine the mineralogies of O-anomalous grains by scanning electron microscopy (SEM; JEOL 6500F) and energy-dispersive X-ray spectroscopy (EDX; Oxford Instruments), using a 5–10 kV accelerating voltage and a 1 nA beam current. This approach works best for relatively large (e.g., >300 nm) grains, as these can be identified more easily in the SEM and there is less contaminating signal from surrounding or underlying phases. If we were unable to exactly locate the grains in the SEM, the mineralogies were estimated from their $Si^-/O^-$ and $AlO^-/O^-$ ratios, although this approach is unreliable (Nguyen et al. 2007, 2010). Based on the high ratios of presolar silicates/oxides determined with Auger spectroscopy in CCs (Floss and Haenecour 2016a), it is likely that the majority of O-anomalous grains identified are silicates rather than oxides.

Contrary to the O-anomalous grains, it is not clear whether all phases with C-anomalous isotopic compositions are older than the Solar System, i.e., “presolar”. SiC and graphite are well-known C-anomalous presolar phases that formed in circumstellar environments (Zinner 2014), whereas it is not yet known whether C-anomalous organic matter formed in the interstellar medium (presolar), in the protoplanetary disk (Floss and Stadermann 2009a; Alexander et al. 2017a), or in both. All C-anomalous grain candidates have a minimum size of nine pixels (e.g., Fig. 1b). SiC has a higher secondary ion yield of $Si^-$ compared to that of silicates, so presolar SiC grains were identified by their C-anomalous isotopic compositions being at least 3.5σ from the matrix means and being associated with $^{28}Si$ in the ion images. Mainstream grains can be roughly distinguished from AB grains and from most C, X, and Y grains by their $^{12}C/^{13}C$ isotopic ratios (Zinner 2014, and references therein). However, reliable classification of SiC grains is not possible without correlated Si and N isotopic data, which are not available for these grains.

All ROIs that were not clearly associated with $^{28}Si$ were defined as C-anomalous grains if their C-isotopic anomaly was at least 4σ at a >0.2 μm ROI diameter, or at least 5σ for smaller

ROIs. These thresholds were chosen on the basis of NanoSIMS image simulations (Supplementary Material). These could be presolar graphite, organic matter, or for very small grains, even extremely $^{13}$C-rich SiC for which isotopic dilution has erased the intrinsic Si signal but not fully hidden the C anomalies. Reliable phase identification is not possible with our dataset. Low $Si^-/C^-$ ratios (e.g., <0.1) and relatively modest C isotopic anomalies (i.e., $\delta^{13}$C in the range of ±300; Alexander et al. 2017a; Nittler et al. 2018a) might be indicative of organic particles (Floss et al. 2004), whereas significantly larger anomalies might be indicative of presolar graphite (Zinner 2014, and references therein). In the following, the term "C-anomalous grains" will refer to all particles that exhibit C isotopic anomalies that fulfill the aforementioned σ-criteria, independent of their phase. Based on the definitions above, we will further distinguish between "SiC grains" (confirmed presolar SiCs by the presence of $^{28}$Si) and "carbonaceous grains" (all other grains, mostly organics; Fig 1c).

The matrix-normalized O- and C-anomalous grain abundances were calculated by dividing the total area covered by these grains by the total area analyzed, excluding by hand in the $^{16}$O image large fragments, cracks, or holes in the thin sections ("corrected area"). Grain sizes were primarily estimated from the NanoSIMS images. In recent work (e.g., Nittler et al 2018a, 2019, 2021) we have assumed that the finite size of the primary ion beam leads to beam-broadening such that presolar grains appear larger than they really are and have applied a simple correction for this effect to all grains. However, as discussed in the Supplementary Material, detailed image simulations reveal this to have been an overly simplistic assumption. Depending on the magnitude and amplitude of a given isotopic anomaly as well as the overall secondary ion count rate, a given grain may appear larger or smaller in NanoSIMS images. Based on realistic image simulations for select grains from the current dataset, we have chosen to only apply the beam broadening correction to grains with an apparent size below 250 nm in the NanoSIMS images. The uncertainties of reported presolar grain abundances are 1σ and were in a few areas (indicated in Table 1) determined by means of the Monte Carlo method of Nittler et al. (2018a). For all other abundances, uncertainties were determined from the confidence limits for small numbers of events provided by Gehrels (1986). Uncertainties calculated with the Monte Carlo method are typically larger and better account for rare, unusually large grains. However, this method only works well for >≈10 grains.

Six cross-sections containing seven presolar grains from Semarkona (see Supplementary Material Table S1) were extracted by focused ion beam (FIB) lift-out and studied with TEM by Singerling et al. (2022).

## 3. Results

To determine accurate presolar grain abundances, we restricted our study to UOCs of very low petrologic types: three thin sections of MET 00526 (L(LL)3.05; Find; MET 00526-15, MET 00526-18 and MET 00526-25), one section of Semarkona (LL3.00; Fall; UNM102, containing two chips; Fig. 2) and one section of NWA 8276 (L3.00; Find; D. Hezel). Matrix areas targeted for NanoSIMS mappings were either chosen after examination by SEM, avoiding clast-rich and/or visibly altered domains, or placed directly adjacent to the TEM sections studied by Dobrică and Brearley (2020a, b). A total corrected area of 26,320 µm$^2$ in three individual matrix areas was mapped in Semarkona, 83,865 µm$^2$ in 18 MET 00526 areas and 29,363 µm$^2$ in 10 NWA 8276 areas. The NanoSIMS maps in one additional MET 00526 area ("AX") are not included in this breakdown. This area was relatively out of focus in the NanoSIMS maps (spatial resolution > 200 nm), preventing us from determining accurate grain sizes for abundance calculations (see Section 3.1.). We excluded these grains from abundance calculations but report their isotopic compositions in Section 3.2.

All Semarkona maps were placed directly adjacent to TEM sections C89, C90A and C90B (Fig. 2, Table 1; Dobrică and Brearley 2020a). These regions are dominated by amorphous silicates that show minimal evidence for secondary processing. The absence of phyllosilicates in these areas further demonstrates that they largely escaped the effects of aqueous alteration seen in other regions of Semarkona (Alexander et al. 1989). In MET 00526, 15 maps were chosen based on SEM data, and two maps were placed directly adjacent to the locations of the TEM sections C28 and C30 (Table 1; Dobrică and Brearley 2020b). MET 00526 is slightly more metamorphosed (petrologic type: 3.05) than Semarkona (3.00). The FeO-rich olivines and phyllosilicates detected in both TEM sections are indicative of an early stage of hydrothermal alteration and fluid-assisted metamorphism in these areas. All NWA 8276 maps were chosen based on SEM observations and contain fayalitic olivines and phyllosilicates. The characteristics and isotopic compositions of all O- and C-anomalous grains reported in the following are listed in the Supplementary Material (Table S1).

### *3.1. Abundances of O- and C-anomalous grains in UOCs*

We detected a total of 122 O-anomalous and 101 C-anomalous grains across all UOCs studied. Of these, 33 O- and 36 C-anomalous grains were identified in Semarkona. Of the C-anomalous grains, 21 are SiC grains and 15 are carbonaceous grains. In MET 00526, we found 78 O- and 46 C-anomalous grains, of which 41 are SiC. MET 00526 area AX contains three additional O-anomalous (M526-01, -02, -03) and nine additional C-anomalous grains (M526-SiC 01–08 and M526-C-01; Supplementary Material Table S1) that were excluded from abundance calculations. NWA 8276 contains 8 O- and 10 C-anomalous grains. Nine grains are SiC and one is a carbonaceous grain. The matrix-normalized abundances of O-anomalous grains with 1σ uncertainties are $151\,^{+50}_{-46}$ ppm in Semarkona, $55\,^{+11}_{-10}$ ppm in MET 00526 and $12\,^{+6}_{-3}$ ppm in NWA 8276 (Table 1, Fig. 3). The matrix-normalized abundances of all C-anomalous grains are $109\,^{+26}_{-22}$ ppm in Semarkona, $24\,^{+5}_{-4}$ ppm in MET 00526 and $16\,^{+7}_{-5}$ ppm in NWA 8276. Excluding carbonaceous grains results in SiC abundances of $53\,^{+14}_{-12}$ in Semarkona, $22\,^{+5}_{-4}$ in MET00526 and $15\,^{+7}_{-5}$ in NWA 8276.

The grain abundances of individual matrix areas in which O-anomalous presolar material is present range from $107\,^{+42}_{-33}$ ppm to $191\,^{+98}_{-89}$ ppm in Semarkona, from $7\,^{+17}_{-6}$ ppm to $165\,^{+111}_{-71}$ ppm in MET 00526 and from $7\,^{+17}_{-6}$ ppm to $49\,^{+67}_{-26}$ in NWA 8276 (Fig. 4). All abundances observed in Semarkona and NWA 8276 areas overlap within 1σ uncertainties with the abundances in other areas of the same sample, whereas this is not the case for several MET 00526 areas. The O-anomalous presolar grain abundance in Semarkona area F2A2 would significantly drop from ~191 ppm to ~62 ppm if two particularly large grains (~1 µm diameter each) had been missed by the randomly placed NanoSIMS grid (cf. Section 3.2.; Fig. 2).

The grain abundances of individual matrix areas in which C-anomalous material was detected range from $84\,^{+29}_{-22}$ ppm to $224\,^{+177}_{-107}$ ppm in Semarkona, from $5\,^{+12}_{-4}$ ppm to $56\,^{+24}_{-17}$ ppm in MET 00526 and from $7\,^{+17}_{-6}$ ppm to $106\,^{+139}_{-68}$ ppm in NWA 8276 (Fig. 4). Most C-anomalous grain abundances overlap within 1σ uncertainties with the abundances in other areas of the same samples. Since most C-anomalous grains in MET 00526 and NWA 8276 are SiC, excluding carbonaceous grains does not significantly decrease the grain abundances reported above. This is not the case for Semarkona where C-anomalous carbonaceous grains are much more abundant. Semarkona SiC abundances in individual areas range from $39\,^{+51}_{-25}$ ppm to $56\,^{+23}_{-17}$ ppm. A relatively large cluster of carbonaceous grains detected in MET 00526 (M526-C-11; Fig. 1c) was excluded from the abundance calculations as its size (~2 µm$^2$ area) would increase the abundance of carbonaceous grains by a factor of 7.

### *3.2. Characteristics and isotopic compositions of O- and C-anomalous grains in UOCs*

The isotopic ratios of O-anomalous presolar grains in all UOC samples range from $8.0\times10^{-4}$ to $3.7\times10^{-3}$ in $^{18}O/^{16}O$ ($\delta^{18}O$ = -599‰ to 862‰) and from $1.9\times10^{-4}$ to $2.9\times10^{-3}$ in $^{17}O/^{16}O$ ($\delta^{17}O$ = -505‰ to 6599‰; Fig. 5; Supplementary Material Table S1). Most O-anomalous grains (102 total, 83.6%) are Group 1 grains (Nittler et al. 1997, 2008; Floss and Haenecour 2016a) with $^{17}O$ enrichments and solar or slightly sub-solar $^{18}O/^{16}O$ ratios. Group 3 (6 grains; 4.9%) and Group 4 grains (11 grains; 9.0%) have relatively low abundances. Only three Group 2 grains (2.5%) have been detected, one in Semarkona and two in MET 00526. However, it is likely that some Group 1 grains are in fact Group 2 grains whose $^{18}O$-depletions have been diluted by surrounding material (cf. Section 4.5. and Supplementary Material; all data are uncorrected for isotopic dilution).

The average diameter of presolar O-anomalous grains is largest in Semarkona (0.33 ±0.20 µm; 1σ uncertainty), intermediate in MET 00526 (0.25 ±0.08 µm) and smallest in NWA 8276 (0.23 ±0.05 µm). The average diameter across all samples is 0.27 ±0.13 µm. Even though we detected more than twice as many presolar O-anomalous grains in MET 00526, the largest grains were found in Semarkona. Area F2A2 contains the two largest grains with diameters of 1.09 µm (SEM-F2-08; Fig. 6a) and 0.90 µm (SEM-F2-15; Fig. 1a). Two more Semarkona grains with diameters of 0.66 µm and 0.65 µm (SEM-F2-29 and SEM-F1-1) are about the same size as the largest MET 00526 grain (0.69 µm; M526-69; Fig 6d). The largest grain detected in NWA 8276 has a diameter of 0.34 µm (N8276-04). All grain sizes are defined as the diameter of a circular grain with the equivalent area to the presolar grain ROI (which is often not circular in shape).

Based on $^{28}Si^-/^{16}O^-$ and $^{27}Al^{16}O^-/^{16}O^-$ secondary ion ratios and, for a few grains, SEM-EDX analyses, we estimate that the majority of O-anomalous grains are silicates (~96%). We detected one presolar Mg-Al spinel in Semarkona and at least five oxides of unknown compositions in MET 00526. The most unusual grain is SEM-F2-08 (Fig. 6a–c). It is the largest grain detected (>1 µm), and it has an amoeboidal shape and a unique mineralogy. It is a complex oxide-silicate composite grain that contains several subgrains of various phases (forsteritic olivine, Mg-Al spinel, Ca-rich pyroxene). The grain's microstructure as well as its elemental and mineralogical compositions will be studied in detail by TEM. Another TEM study by Singerling et al. (2022) investigated the characteristics and compositions of six other presolar grains from Semarkona (1 spinel, 4 silicates and 1 SiC; see Supplementary Material Table S1). M526-69 is another large (0.69 µm diameter; Fig. 6 d–i) composite grain. In the ion image, the upper subgrain is clearly enriched in Al ($AlO^-/O^-$ = 0.04) but relatively low in Si ($Si^-/O^-$ = 0.01), whereas the opposite is seen in the lower subgrain ($AlO^-/O^-$ = 0.005 and $Si^-/O^-$ = 0.03). An EDX element map (Fig. 6f) confirms the presence of Al and Ti in the upper subgrain (likely corundum), while the rest of the grain is dominated by Ca (possibly diopside). Both subgrains have similar isotopic anomalies in $\delta^{17}O$ (~1000‰) and $\delta^{18}O$ (~ -500‰).

The $^{12}C/^{13}C$ ratios of all SiC grains detected in this study range from ~3 to 182 ($\delta^{13}C \approx$ 27,000 to -500‰, respectively). A total of 65 grains (82%) with $^{12}C/^{13}C$ ratios between 10 and 100 are probably mainstream grains (Zinner 2014, and references therein). One grain from Semarkona, six grains from MET 00526, and three NWA 8276 grains (10 total, 13%) are AB grains with $^{12}C/^{13}C$ ratios <10. Four grains in MET 00526 (5%) have $^{12}C/^{13}C$ ratios between 137 and 182 and are most likely X, Y, or C grains. No similar $^{13}C$-depleted grains were found in Semarkona or NWA 8276. The average diameter of SiC grains with 1σ uncertainties in all UOCs studied is 0.25 ±0.07 µm, and within error the same in all samples. The diameters of the largest SiC grains in MET 00526 (0.40 µm) and Semarkona (0.38 µm) are similar. The largest grain in NWA 8276 has a 0.31 µm diameter.

Most C-anomalous carbonaceous grains have $^{12}C/^{13}C$ ratios in the range of ~70–130 (i.e., $\delta^{13}C$ = +300‰ to -300‰), which is typical for organic matter (e.g., Nittler et al. 2018a). One outlier is depleted in $^{13}C$ ($^{12}C/^{13}C \approx 212$; $\delta^{13}C$ = -579‰) and one outlier is enriched in $^{13}C$ ($^{12}C/^{13}C \approx 56$; $\delta^{13}C$ = 579‰, which is similar to the average composition of mainstream SiC grains). Another outlier, N8276-C-01, is very $^{13}C$-rich ($^{12}C/^{13}C \approx 12$; $\delta^{13}C$ = 6753‰) and might be a presolar graphite (Zinner 2014, and references therein). The average grain diameter of all carbonaceous grains is 0.29 ±0.14 µm. The largest grain was found in Semarkona and has a diameter of 0.42 µm. Grain SEM-C-01 is a clump of three $^{13}C$-poor grains with a combined diameter of 0.77 µm. Another large cluster of $^{13}C$-poor carbonaceous grains (~2 µm$^2$ area, M526-C-07; Fig. 1c) was found in MET 00526. It consists of at least 10 grains that appear loosely fused together, mostly fulfill the σ-criterion (cf. Section 2), have $\delta^{13}C$ anomalies of approximately -200‰, and in parts show some overlap with Si. Additional measurements (e.g., H and N isotopes, Raman spectroscopy) could provide clues about the origin of this unusual cluster of C-rich grains.

## 4. Discussion

### *4.1. Presolar O-anomalous grain abundances in UOCs.*

The presolar oxide grain inventory of UOCs has been well-studied in grain separates from acid residues (Nittler et al. 1994, 2008; Choi et al. 1998; Zinner et al. 2005). Accurate abundances of O-anomalous presolar grains cannot be determined from acid residues as presolar silicates, the most common demonstrably presolar phase in astromaterials (Floss and Haenecour 2016a), are destroyed during sample preparation. To our knowledge, only four presolar silicates in UOCs have been reported in the published literature: one grain from Krymka (Hoppe et al. 2018) and three grains from Semarkona (Hoppe et al. 2021). The only previous work reporting other presolar silicates (~90 grains total) and/or attempting to calculate O-anomalous grain abundances in UOCs are unpublished abstracts (Mostefaoui et al. 2003, 2004; Tonotani et al. 2006; Leitner et al. 2014; Floss and Haenecour 2016b, c). This makes our study the first comprehensive in situ investigation of O-anomalous presolar grain abundances in UOCs. Nonetheless, we cautiously compare our data to unpublished reports, as most of these primarily used the same methodology (except Tonotani et al. 2006), as well as fairly standard protocols for data reduction.

We found substantially higher abundances of O-anomalous presolar grains in Semarkona than in previous studies ($151^{+50}_{-46}$ ppm vs. 4–15 ppm; Mostefaoui et al. 2003; Tonotani et al. 2006; no abundance was reported by Hoppe et al. 2021 for their Semarkona grains; Fig. 3). Since our maps were placed in relatively pristine matrix areas, our data probably provide the best possible estimate of the pre-alteration abundance of O-anomalous grains in Semarkona. However, our inferred abundances might still be lower than the original presolar grain abundances in Semarkona due to potential destruction of delicate grains during low degrees of parent body alteration. Additionally, or alternatively, pre-accretionary alteration could have potentially destroyed some presolar grains in Semarkona's formation region (Haenecour et al. 2018; Dobrică and Brearley 2020a). Contrary to Semarkona, NWA 8276 matrix and all matrix areas studied by Mostefaoui et al. (2003) and Tonotani et al. (2006) in several other UOCs (i.e., Bishunpur, Krymka, Jiddat al Harasis 026 and Dhofar 008) have abundances <15 ppm. The presolar grain abundances seem to increase when more pristine areas of the same chondrites are targeted, such as fine-grained rims of cluster chondrite clasts in Krymka (~92 ppm; Leitner et al. 2014). The often very low abundances of presolar O-anomalous grains in UOCs, thus, almost certainly reflect destruction of presolar material during secondary processing, rather than significantly lower abundances of presolar grains in the formation regions of UOCs compared to CCs. It is challenging to discriminate between the extent of secondary processing that occurred on meteorite parent bodies versus that which occurred after the meteorite's fall on Earth. Meteorites such as NWA 8276 and MET 00526 that were found after residing in hot or cold deserts could have been modified by terrestrial processes. In NWA 8276, terrestrial weathering might have largely destroyed presolar O-anomalous grains, explaining the very low abundances. As a meteorite fall, Semarkona probably escaped terrestrial weathering. The low presolar grain abundances in aqueously altered Semarkona matrix (Mostefaoui et al. 2003; Tonotani et al. 2006), compared to high abundances in relatively pristine matrix areas (this study), show that parent body processes are perfectly capable of producing major differences in presolar O-anomalous grain abundances within and between chondrites.

Floss and Haenecour (2016b) reported an extraordinarily high O-anomalous presolar grain abundance of ~275 ±50 ppm in MET 00526, higher even than the highest abundances seen in the most primitive CCs (Fig. 7, and references therein). Our O-anomalous grain abundance for MET 00526 is significantly lower ($55\,^{+11}_{-10}$ ppm; Fig. 3). However, we mapped a much larger area (83,865 vs. 19,200 µm$^2$), split over three different thin sections and a total of 18 separate matrix areas. The better overall statistics allows for a much more detailed and representative assessment of MET 00526's presolar grain inventory. Our data are thus less likely to be biased by local heterogeneities in the matrix, or by detection of rare, particularly large presolar grains (>1 µm; see F2A2 in Fig. 4). Floss and Haenecour (2016b) did not report presolar grain sizes in their abstract. Their grain abundance was determined from 32 O-anomalous presolar grains found in an area of 19,200 µm$^2$, implying an average grain diameter of ≈460 nm. The average grain sizes determined in the present study and in most studies of presolar silicate grains in CCs are significantly smaller (typically <300 nm). We note that Floss and Haenecour (2016b) used the L'image software for data reduction instead of the custom software used in other presolar grain studies from the same group (cf. Stadermann et al. 2005) and did not provide details as to how ion images were processed. Thus, it is possible that the large average grain size (and hence high abundance) reported in their abstract is an artifact of their image processing methodology. If Floss and Haenecour (2016b) indeed overestimated their grain sizes and the true average size of their presolar grains was between 250 and 300 nm, then their inferred abundance would drop to 80–100 ppm, which is within the range of the present study.

Assuming that the three (out of 18) individual MET 00526 areas of the present study with the highest abundances are the most pristine (cf. Fig. 4; areas A05, A12, B03), the best estimate of the O-anomalous presolar grain abundance in MET 00526 would be ≥100–160 ppm, or higher. This abundance would be consistent with the lower limits of O-anomalous presolar grain abundances in the relatively pristine Semarkona matrix ($151\,^{+50}_{-46}$ ppm) and the ~145 ±30 ppm reported for Queen Alexandra Range (QUE) 97008 (L3.05; Floss and Haenecour 2016c). The abundances in most primitive CCs are the same within error, although the least altered CCs seem to have slightly higher abundances (see Fig. 7; up to 240 ppm in DOM 08006 and Asuka 12169; Haenecour et al. 2018; Nittler et al. 2018, 2021).

*4.2. Presolar C-anomalous grain abundances in UOCs.*

The reported abundances of presolar SiC grains in the CCs typically range from ~10 ppm to ~60 ppm (cf. Fig. 7) but can occasionally be as high as 182 ±33 ppm (in Grove Mountain 021710; Zhao et al. 2013). Abundances found in samples recently returned from carbonaceous-type asteroid Ryugu by the Hayabusa2 spacecraft fall within the same range ($25^{+6}_{-5}$ ppm; Barosch et al. 2022). The matrix-normalized SiC abundances of several UOCs have been estimated from noble gas analyses in acid-resistant residues (Ne-E(H); Huss and Lewis 1995) and range from <1 ppm to ~20 ppm. Only two of ten UOCs studied – Semarkona and Bishunpur – have SiC abundances that are estimated to be higher than 10 ppm. SiC abundances based on in situ SIMS analyses of UOCs reported in various abstracts range from <1 ppm (Tonotani et al. 2006) to 125 ±30 ppm (Floss and Haenecour 2016c). However, Semarkona is the only UOC in which a systematic NanoSIMS search for SiC grains has been conducted and published (Davidson et al. 2014), albeit in an organics-rich acid residue. These authors found an abundance of $30^{+15}_{-10}$ ppm, comparable within error to an estimate of ~20 ppm by Huss and Lewis (1995). Our SiC abundance, determined from in situ mapping of relatively pristine Semarkona matrix, is $53^{+14}_{-12}$ ppm. If this represents the original abundance of presolar SiC grains in Semarkona, the abundances in UOC matrices and most primitive CC matrices appear to be quite similar (cf. Davidson et al. 2014; Haenecour et al. 2018).

In MET 00526 and NWA 8276, we found relatively low abundances of presolar SiC grains ($22^{+5}_{-4}$ ppm and $15^{+7}_{-5}$ ppm, respectively; Table 1, Fig. 3). In both meteorites, the individual areas with the highest abundances of presolar O-anomalous grains are different from the areas with the highest abundances of SiC grains. It is particularly puzzling that MET 00526 areas A05 and B03 are among the three "pristine" areas with the highest presolar O-anomalous grain abundances (Table 1, Fig. 4), and simultaneously among the three areas with the lowest SiC abundances. This is most likely a statistical fluke. Presolar silicates are the dominant presolar phase (Floss and Haenecour 2016a) and the comparatively low SiC abundances thus have much larger uncertainties due to poor counting statistics. There is no indication that any secondary process would destroy SiC grains preferentially to O-anomalous grains. Relative to presolar silicates, SiC grains are more resistant to parent body metamorphism and aqueous alteration (Huss et al. 2003; Davidson et al. 2014; Floss and Haenecour 2016a; Haenecour et al. 2018). Low abundances of presolar SiC are typically seen in meteorites that experienced significant heating (Huss and Lewis 1995) and/or prolonged periods of aqueous alteration in the presence of an oxidant (Davidson et al. 2014). Our samples are of low petrologic types and did not experience significant heating (Alexander et al. 1989; Dobrică and Brearley 2020a, b). However, some SiC grains might have been modified and/or destroyed during prolonged aqueous alteration (Riebe et al. 2020), particularly in NWA 8276.

Many previous in-situ studies of presolar grains either do not report C-anomalous carbonaceous grain abundances, or they do not distinguish these from SiC abundances. Carbonaceous grains cannot be distinguished into different phases (i.e., graphite, SiC, organics) based on C-isotopic data alone, which makes comparing abundances in different chondrites problematic. The majority are probably the C-anomalous organic matter discussed in Section 4.4. In CCs, carbonaceous grain abundances are variable, e.g., $9\,^{+5}_{-4}$ ppm in CI chondrites (Barosch et al. 2022), $\sim 33$ ppm in CO chondrites (Nittler et al. 2018a) and up to ~120 ±23 ppm in CR chondrites (Floss and Stadermann 2009). Only Semarkona has relatively high abundances of carbonaceous grains ($56\,^{+19}_{-14}$ ppm; Table 1).

*4.3. Composition and distribution of presolar dust across major chondrite-forming reservoirs.*

Over the past decade, a growing number of studies have reported evidence for an isotopic dichotomy in the Solar System, most prominently in the bulk meteorite isotopic compositions of Cr, Ti, Ni and Mo (Warren 2011; Kruijer et al. 2019, and references therein). Although still debated, these isotopic differences are largely interpreted as an inner-outer Solar System dichotomy with CCs (and isotopically related achondrites and iron meteorites) having formed outside Jupiter's orbit, while the NC ("non-carbonaceous," including OCs, Rumuruti, and enstatite chondrites, Earth, Mars, etc.) reservoir formed in the inner Solar System. The bulk isotopic differences between the CC and NC reservoirs may reflect a heterogenous distribution and subsequent incorporation of presolar material from various stellar sources as carriers of nucleosynthetic isotope anomalies, as well as limited exchange of material between chondrite-forming reservoirs in the outer (CC) and inner (NC) protoplanetary disk. Even though we did not study the aforementioned isotope systems in which bulk meteorite differences are most prominent, the types, morphologies, abundances, and/or isotopic compositions of presolar grains in the most primitive members of CC and NC bodies might be different. Our new dataset allows a first in situ comparison of the characteristics and distribution of presolar dust across the major chondrite-forming reservoirs in the protoplanetary disk.

There are no obvious differences in the matrix-normalized abundances of O- and C-anomalous presolar dust in relatively pristine UOC and lightly altered CC matrices (cf. Section 4.1. and 4.2.; Fig. 7). This similarity might indicate a relatively homogenous distribution of presolar dust across chondrite-forming reservoirs in the protoplanetary disk (cf. Huss and Lewis 1995; Huss et al. 2003; Davidson et al. 2014), even though the formation times of chondrite groups probably differ by up to ~2 Ma (e.g., Sugiura and Fujiya 2014; Desch et al. 2018). Small-scale heterogeneities within these reservoirs (e.g., Wasson and Rubin 2009) and/or locally different degrees of secondary processing (cf. Section 4.1.), could then be responsible for the variations of presolar grain abundances seen within the matrices of their respective chondrite hosts. This is most obvious in Semarkona, where aqueously altered matrix areas have much lower abundances of presolar (silicate) grains (Mostefaoui et al. 2003; Tonotani et al. 2006) than their more pristine counterparts (this study; Section 4.1.). We note that the O-anomalous grain abundances in all chondrites are apparently lower compared to cometary samples including interplanetary dust particles, some Antarctic micrometeorites, and comet Wild 2 samples returned by NASA's Stardust spacecraft (Floss and Haenecour 2016a, and references therein; Alexander et al. 2017b).

The isotopic compositions of the presolar O-anomalous grains span similar ranges to most grains identified in situ in primitive CCs (Fig. 5, and references therein). Occasionally, grains with more extreme O isotopic compositions have been identified in CCs that have not yet been found in UOCs (e.g., "extreme" Group 1 grains with $^{17}O/^{16}O > 3\times10^{-3}$; Nguyen et al. 2007; Vollmer et al. 2008). The fractions of Group 1–4 O-anomalous presolar grains accreted by inner and outer Solar System chondrites appear to be similar. To exclude analytical biases, only grains identified in situ with a NanoSIMS are included in this breakdown: Group 2 (2% each) and Group 4 (11% vs. 9%) fractions are almost identical in UOCs and CCs, respectively, and $^{17}O$-rich Group 1 fractions (84% vs. 86%) and Group 3 fractions (5% vs. 2%) are only slightly different (CC group fractions were taken from Floss and Haenecour 2016a, and references therein). The observation that O-anomalous grains in UOCs and CCs have similar ranges of O-isotopic compositions and accreted an almost identical mix of Group 1–4 grains indicates that largely the same O-anomalous presolar dust was present in both major chondrite-forming reservoirs.

It is particularly interesting that Group 4 ($^{18}O$-enriched) grain fractions are equal in primitive UOCs and CCs. Group 4 grains are thought to have primarily formed in Type II supernovae (Choi et al. 1998, Nittler et al. 2008) and it has been suggested that most Group 4 oxide grains may have originated from a single supernova event that injected the protoplanetary disk with $^{18}O$-rich material (Nittler et al. 2008). Yada et al. (2008) speculated that asteroids that formed in different locations in the solar nebula and/or at different times may have incorporated different proportions of Group 4 grains. This could explain the relatively lower Group 4 grain abundances in CCs (~9%) compared to Antarctic micrometeorites and interplanetary dust particles (24–31%; Floss and Haenecour 2016a, and references therein), which may sample a cometary reservoir that formed substantially farther from the Sun than chondrites. However, the similar Group 4 fractions in CCs and UOCs do not support this as an explanation of the CC-NC isotopic dichotomy indicated by chondrites.

We report remarkable similarities in the presolar grain populations of meteorites from outer and inner Solar System reservoirs and did not find definitive evidence for mineralogical or isotopic differences. However, the dataset obtained so far is insufficient to detect the relatively small variations in presolar grain abundances and/or the differences in their isotopic compositions that may be needed to produce the differences in bulk nucleosynthetic anomalies that characterize the NC-CC dichotomy. As mentioned above, none of the isotopes with prominent bulk meteorite NC-CC isotopic differences have been studied here (cf. Kruijer et al. 2019). Furthermore, we only identified and discussed presolar grains with sizes >130 nm. Presolar grains smaller than 100 nm were missed as these can only be detected in higher-resolution NanoSIMS surveys as shown by Hoppe et al. (2015, 2017) for CR2 and C2-ungr. chondrites. The authors further suggested that O-rich supernova grains, potential carriers of neutron-rich isotopes, might be more abundant at smaller sizes. Nittler et al. (2018b) studied an acid residue of the CI chondrite Orgueil and detected the most extreme $^{54}$Cr-enrichments in presolar grains with diameters <80 nm (see also Dauphas et al. 2010; Qin et al. 2011). Therefore, expanding our dataset by remeasuring UOC presolar grains for isotopes of, e.g., Mg, Si, Ti, and Cr, and including high-resolution ion mappings of UOCs will be highly beneficial to better examine and compare presolar dust in chondrites from different Solar System reservoirs.

### *4.4. Organic matter in UOCs*

The dominant type of organic matter in primitive meteorites is a macromolecular material that is impervious to traditional solvents and acids used for demineralization (insoluble organic matter or IOM). The isotopic composition of this material has been mostly studied in acid residues of UOCs and CCs (e.g., Alexander et al. 1996, 2007, 2017a, and references therein; Remusat et al. 2016, 2019), and occasionally in situ with the NanoSIMS (not yet in UOCs, e.g., Floss and Stadermann 2009a; Bose et al. 2012; Piani et al. 2012; Floss et al. 2014; Le Guillou et al. 2014; Nittler et al. 2018a). It is still debated how and where IOM or its precursor materials were formed (Alexander et al. 2017a; Piani et al. 2021), whether all chondrites accreted the same (Alexander et al. 2007) or different organics (Remusat et al. 2016), and to what degree these were subsequently modified by secondary processes on chondrite parent bodies. IOM is known to be isotopically diverse on a sub-μm to whole-rock scale, often showing $^{15}N$ and/or D isotope enrichments and/or depletions. Most C-anomalous carbonaceous grains reported here are probably IOM. The range of C isotopic compositions ($\delta^{13}C$ from 300‰ to -300‰) in UOC carbonaceous grains is similar to IOM from primitive CCs (Floss and Stadermann 2009a; Bose et al. 2012; Alexander et al. 2017a; Nittler et al. 2018a), indicating a similar C isotopic compositional diversity in both chondrite classes. This might suggest that both chondrite classes accreted IOM or its precursor material(s) from the same source(s). Determining H and N isotopic compositions will be necessary to better compare organic matter in UOCs and CCs.

Only a small fraction of IOM in primitive CCs (<5%; Floss and Stadermann 2009a; Nittler et al. 2018a) exhibits C isotope anomalies. Thus, the carbonaceous grain abundances reported here (Table 1) do not represent total IOM contents in UOCs. Alexander et al. (2007) determined bulk insoluble C contents of 0.36 wt.% for Semarkona and 0.21 wt.% for MET 00526. Our results indicate that carbonaceous grains are common constituents of relatively pristine UOC matrices. Abundant submicron C-anomalous carbonaceous grains (15 total, $56^{+19}_{-14}$ ppm; Table 1, Fig. 3) are dispersed throughout the matrix of Semarkona. A much smaller number of such grains (6 total, <3 ppm) was detected in a much larger total matrix area mapped in MET 00526 and NWA 8276. The lower abundances of carbonaceous grains in MET 00526 matrix compared to Semarkona may be in part explained by the lower bulk insoluble C contents in MET 00526 (see above; Alexander et al. 2007). Furthermore, the higher degree of fluid-assisted metamorphism in MET 00526 (L/LL3.05) matrix compared to the relatively pristine Semarkona (LL3.00) matrix areas studied here (Dobrică and Brearley 2020a, b) is likely to be responsible for the observed differences. As for presolar grains, terrestrial weathering could have also destroyed organic material, particularly in NWA 8276. The evolution of IOM during secondary aqueous alteration is poorly understood and the degree to which this material was modified is controversial (e.g., Alexander et al. 2007, 2017a; Remusat et al. 2016). To elucidate the evolution of organic material(s) during aqueous alteration, a detailed investigation and comparison of organic matter in relatively pristine and more altered regions of Semarkona matrix might be highly beneficial.

Secondary processes may also be responsible for local accumulation of carbonaceous grains, forming irregular clusters such as SEM-C-01 and M526-C-07 (Fig. 1c), nanoglobules, or veins (Le Guillou et al. 2014; Alexander et al. 2017a). These probably reflect material redistribution during melting of ices accreted by chondrite matrices. Large, irregular aggregates similar to M526-C-07 have been found in primitive CCs (e.g., see Fig. 3 in Floss et al. 2014), indicating that carbonaceous grains exhibit the same variety of morphologies in both chondrite clans. We note that the H, C, N isotopic compositions of carbonaceous grains are not correlated with grain morphologies and can be highly diverse on a micron-scale (Matrajt et al. 2012; Floss et al. 2014; Alexander et al. 2017a).

*4.5. NanoSIMS image simulations*

The image simulations presented in this study helped to better understand the effects of primary ion beam tailing and counting statistics on the identification and quantification of presolar grains. A detailed description of methods and results can be found in the Supplementary Material.

We have found that the degree of isotopic dilution for a given isotopically anomalous region depends not only on the relative size of the region and primary beam, but also significantly on the magnitude and direction (enrichment or depletion) of the anomaly and the overall signal strength (e.g., counting statistics), precluding accurate corrections for dilution. Some previous studies reported presolar grain compositions that were corrected for isotope dilution, however, details about the procedures used were not always included. To better compare datasets, it would be beneficial to always report uncorrected data in addition to corrected data.

All data reported in the present work are uncorrected for isotopic dilution. Therefore, some grains may have been misclassified and the reported isotopic anomalies for most grains should be considered as lower limits. For example, some O-anomalous Group 2 grains could have been misclassified as Group 1 grains. Our results also indicate that presolar grain sizes cannot always be determined accurately, implying additional uncertainty in the abundance estimates beyond Poisson statistics.

We determined new significance thresholds for the anomalies of grain candidates, reducing the risk of including false candidates. The reported criteria (Section 2) will help to present internally consistent datasets in future studies, correct previously reported data that relied on less stringent thresholds, and may ease the comparison of data from other working groups that use different criteria.

## 5. Conclusions

We have conducted a NanoSIMS-based search for presolar O- and C-anomalous grains in UOCs and present the first detailed in situ investigation of their presolar grain populations. As previously pointed out by Floss and Haenecour (2016b), and confirmed in this study, chondrites of the same low petrologic type (3.00) may still have vastly different presolar grain abundances in their matrices. Presolar silicates are particularly sensitive to metamorphism and aqueous alteration, even at early stages as seen in MET 00526 (Dobrică and Brearley 2020b). Only the most pristine matrices may be able to preserve presolar silicates in abundances close to their original abundances. In contrast to previous (mainly unpublished) reports, we specifically targeted recently identified matrix areas in Semarkona that largely escaped secondary processing (Dobrică and Brearley 2020a) and observed significantly higher presolar grain abundances. Our results indicate that the differences in presolar O-anomalous grain abundances within and between even the most primitive UOCs (i.e., petrologic type 3.00–3.05) do not necessarily result from different initial abundances of presolar dust in nebular reservoirs. Instead, the main mechanism producing differences between and within samples are secondary processes. The slight difference in the degree of parent body aqueous alteration and metamorphism between Semarkona and MET 00526 probably explains the lower O-anomalous presolar grain abundances in MET 00526. Particularly NWA 8276 could have been further modified by terrestrial weathering which probably resulted in the almost complete destruction of its O-anomalous presolar grain population.

Our results emphasize the need to identify the most pristine chondrites to study their presolar grain abundances. Unfortunately, the current classification of meteorites according to their petrologic type does not provide adequate criteria to distinguish samples on the level required to investigate their most delicate components. Presolar O-anomalous grain abundances could potentially serve as an additional criterion for petrologic type classification, although this may be challenging to implement.

Our new dataset also allowed a first comparison of the abundances and characteristics of presolar grains accreted by chondrites that likely formed in different locations of the Solar System and at different times (e.g., Sugiura and Fujiya 2014; Desch et al. 2018). We showed that the most pristine UOC matrices contain similar presolar grain abundances to lightly altered CCs, and that they have comparable isotopic compositions. While presolar grain morphologies, sizes, compositions and abundances seem to be similar in CCs and UOCs, several key aspects remain to be investigated in future studies – e.g., high-resolution NanoSIMS mappings examining the inventory of extremely small presolar grains (<100 nm; Hoppe et al. 2015, 2017) that were missed in this study. Furthermore, studying additional isotopic compositions of UOC presolar grains, such as $^{54}Cr$ and $^{50}Ti$ (cf. Dauphas et al. 2010; Qin et al. 2011; Nittler et al. 2018b), will be necessary to better compare presolar grain populations with regard to CC-NC nucleosynthetic differences. Most grains in the current dataset could also be remeasured for Mg and Si isotopes and compared to the dataset by Hoppe et al. (2021).

## Acknowledgements

We are grateful to the JSC curators and the Meteorite Working Group for the loan of the MET 00526 sections. This work was supported by NASA Emerging Worlds Grant 80NSSC20K0340 (LRN). We thank Smail Mostefaoui, an anonymous referee and associate editor Yves Marrocchi for helpful comments.

## Appendix A. Supplementary Material

The Supplementary Material contains a Table listing the characteristics of all presolar O- and C-anomalous grains identified in this study (Table S1). Furthermore, NanoSIMS image simulations are presented to examine the effects of primary ion beam tailing and counting statistics on the identification and quantification of presolar grains. The results were used to determine significance thresholds for the anomalies of presolar grain candidates (cf. Section 2).

## Footnote A:

$\delta^{13}C = [(^{13}C/^{12}C)_{grain}/(^{13}C/^{12}C)_{std} - 1\,] \times 1000$; $(^{13}C/^{12}C)_{std} = 0.011237$ (VPDB)

$\delta^{x}O = [(^{x}O/^{16}O)_{grain}/(^{x}O/^{16}O)_{std} - 1\,] \times 1000$; $x = 17, 18$; $(^{17}O/^{16}O)_{std} = 0.0003829$, $(^{18}O/^{16}O)_{std} = 0.0020052$ (VSMOW)

## References

Alexander C. M. O'D., Barber D. J. and Hutchison R. (1989) The microstructure of Semarkona and Bishunpur. *Geochim. Cosmochim. Acta* **53**, 3045–3057.

Alexander C. M. O'D., Arden J. W., Ash R. D. and Pillinger C. T. (1990) Presolar component in the ordinary chondrites. *Earth Planet. Sci. Lett.* **99**, 220–229.

Alexander C. M. O'D. (1993) Presolar SiC in chondrites: How variable and how many sources? *Geochim. Cosmochim. Acta* **57**, 2869–2888.

Alexander C. M. O'D., Russell S. S., Arden J. W., Ash R. D., Grady M. M. and Pillinger C. T. (1996) The origin of chondritic macromolecular organic matter: A carbon and nitrogen isotope study. *Meteorit. Planet. Sci.* **33**, 603–622.

Alexander C. M. O'D., Fogel M., Yabuta H. and Cody G. D. (2007) The origin and evolution of chondrites recorded in the elemental and isotopic compositions of their macromolecular organic matter. *Geochim. Cosmochim. Acta* **71**, 4380–4403.

Alexander C. M. O'D., Cody G. D., De Gregorio B. T., Nittler L. R. and Stroud R. M. (2017a) The nature, origin and modification of insoluble organic matter in chondrites, the major source of Earth's C and N. *Chem. Erde* **77**, 227–256.

Alexander C. M. O'D., Nittler L. R., Davidson J. and Ciesla F. J. (2017b) Measuring the level of interstellar inheritance in the solar protoplanetary disk. *Meteorit. Planet. Sci.* **52**, 1797–1821.

Barosch J., Nittler L. R., Wang J., Alexander C. M. O'D., De Gregorio B. T., Engrand C., Kebukawa Y., Nagashima K., Stroud R. M., Yabuta H., Abe Y., Aléon J., Amari S., Amelin Y., Bajo K., Bejach L., Bizzarro M., Bonal L., Bouvier A., Carlson R. W., Chaussidon M., Choi B.-G., Cody G. D., Dartois E., Dauphas N., Davis A. M., Dazzi A., Deniset-Besseau A., Di Rocco T., Duprat J., Fujiya W., Fukai R., Gautam I., Haba M. K., Hashiguchi M., Hibiya Y., Hidaka H., Homma H., Hoppe P., Huss G. R., Ichida K., Iizuka T., Ireland T. R., Ishikawa A., Ito M., Itoh S., Kamide K., Kawasaki N., Kilcoyne D. A. L., Kita N. T., Kitajima K., Kleine T., Komatani S., Komatsu M., Krot A. N., Liu M.-C., Martins Z., Masuda Y., Mathurin J., McKeegan K. D., Montagnac G., Morita M., Mostefaoui S., Motomura K., Moynier F., Nakai I., Nguyen A. N., Ohigashi T., Okumura T., Onose M., Pack A., Park C., Piani L., Qin L., Quirico E., Remusat L., Russell S. S., Sakamoto N., Sandford S. A., Schönbächler M., Shigenaka M., Suga H., Tafla L., Takahashi Y., Takeichi Y., Tamenori Y., Tang H., Terada K., Terada Y., Usui T., Verdier-Paoletti M., Wada S., Wadhwa M., Wakabayashi D., Walker R. J., Yamashita K., Yamashita S., Yin Q.-Z., Yokoyama T., Yoneda S., Young E. D., Yui H., Zhang A.-C., Abe M., Miyazaki A., Nakato A., Nakazawa S., Nishimura M., Okada T., Saiki T., Tanaka S., Terui F., Tsuda Y., Watanabe S., Yada T., Yogata K., Yoshikawa M., Nakamura T., Naraoka H., Noguchi T., Okazaki R., Sakamoto K., Tachibana S., Yurimoto H. (2022) Presolar stardust in asteroid Ryugu. *Astrophys. J. Lett.* **935**, L3.

Bose M., Floss C., Stadermann F. J., Stroud R. M. and Speck A. K. (2012) Circumstellar and interstellar material in the CO3 chondrite ALHA 77307: An isotopic and elemental investigation. *Geochim. Cosmochim. Acta* **93**, 77–101.

Boujibar A., Howell S., Zhang S., Hystad G., Prabhu A., Liu N., Stephan T., Narkar S., Eleish A., Morrison S. M., Hazen R. M. and Nittler L. R. (2021) Cluster analysis of presolar silicon carbide grains: evaluation if their classification and astrophysical implications. *Astrophys. J. Lett.* **907**, L39.

Choi B.-G., Huss G. R., Wasserburg G. J. and Gallino R. (1998) Presolar corundum and spinel in ordinary chondrites: Origins from AGB stars and a supernova. *Science* **282**, 1284–1289.

Dauphas N., Remusat L., Chen J. H., Roskosz M., Papanastassiou D. A., Stodolna J., Guan Y., Ma C. and Eiler J. M. (2010) Neutron-rich chromium isotope anomalies in supernova dust. *Astrophys. J.* **720**, 1577–1591.

Davidson J., Busemann H., Nittler L. R., Alexander C. M. O'D., Orthous-Daunay F-R., Franchi I. A. and Hoppe P. (2014) Abundances of presolar silicon carbide grains in primitive meteorites determined by NanoSIMS. *Geochim. Cosmochim. Acta* **139**, 248–266.

Desch S. J., Kalyaan A. and Alexander C. M. O'D. (2018) The effect of Jupiter's formation on the distribution of refractory elements and inclusions in meteorites. *Astrophys. J. Suppl. Ser.* **238**, 11–42.

Dobrică E. and Brearley A. J. (2020a) Amorphous silicates in the matrix of Semarkona: The first evidence for the localized preservation of pristine materials in the most unequilibrated ordinary chondrites. *Meteorit. Planet. Sci.* **55**, 649–668.

Dobrică E. and Brearley A. J. (2020b) Iron-rich olivine in the unequilibrated ordinary chondrite, MET 00526: Earliest stages of formation. *Meteorit. Planet. Sci.* **55**, 2652–2669.

Floss C., Stadermann F. J., Bradley J., Dai Z. R., Bajt S. and Graham G. (2004) Carbon and nitrogen isotopic anomalies in an anhydrous interplanetary dust particle. *Science* **303**, 1355–1358.

Floss C. and Stadermann F. J. (2009a) High abundances of circumstellar and interstellar C-anomalous phases in the primitive CR3 chondrites QUE 99177 and MET 0426. *Astrophys. J.* **907**, 1242–1255.

Floss C. and Stadermann F. J. (2009b) Auger nanoprobe analysis of presolar ferromagnesian silicate grains from primitive CR chondrites QUE 99177 and MET 00426. *Geochim. Cosmochim. Acta* **73**, 2415–2440.

Floss C. and Stadermann F. J. (2012) Presolar silicate and oxide abundances and compositions in the ungrouped carbonaceous chondrite Adelaide and the K chondrite Kakangari: The effects of secondary processing. *Meteorit. Planet. Sci.* **47**, 992–1009.

Floss C., Le Guillou C. and Brearley A. J. (2014) Coordinated NanoSIMS and FIB-TEM analyses of organic matter and associated matrix materials in CR3 chondrites. *Geochim. Cosmochim. Acta* **139**, 1–25.

Floss C. and Haenecour P. (2016a) Presolar silicate grains: Abundances, isotopic and elemental compositions, and the effects of secondary processing. *Geochem. J.* **50**, 3–25.

Floss C. and Haenecour P. (2016b) Meteorite Hills (MET) 00526: An unequilibrated ordinary chonderite with high presolar grain abundances. *Lunar Planet. Sci. Conf.* **47**, abstr. #6015.

Floss C. and Haenecour P. (2016c) Presolar silicate abundances in unequilibrated ordinary chondrites Meteorite Hills 00526 and Queen Alexandra Range 97008. *79th Annual Meeting of the Meteoritical Society*, abstr. #6015.

Gehrels N. (1986) Confidence limits for small numbers of events in astrophysical data. *Astrophys. J.* **303**, 336–346.

Gyngard F., Zinner E., Nittler L. R., Morgand A., Stadermann F. J. and Hynes K. M. (2010) Automated NanoSIMS measurements of spinel stardust from the Murray meteorite. *Astrophys. J.* **717**, 107–120.

Haenecour P., Floss C., Zega T. J., Croat T. K., Wang A., Jolliff B. L. and Carpenter P. (2018) Presolar silicates in the matrix and fine-grained rims around chondrules in primitive CO3.0 chondrites: Evidence for pre-accretionary aqueous alteration of the rims in the solar nebula. *Geochim. Cosmochim. Acta* **221**, 379–405.

Hoppe P., Leitner J. and Kodolányi J. (2015) New constraints on the abundances of silicate and oxide stardust from supernovae in the Acfer 094 meteorite. *Astrophys. J.* **808**, 9–15.

Hoppe P., Leitner J. and Kodolányi J. (2017) The stardust abundance in the local interstellar cloud at the birth of the Solar System. *Nature Astron.* **1**, 617–620.

Hoppe P., Leitner J. and Kodolányi J. (2018) New insights into the galactic chemical evolution of Magnesium and Silicon isotopes from studies of silicate stardust. *Astrophys. J.* **869**, 47–60.

Hoppe P., Leitner J., Kodolányi J. and Vollmer C. (2021) Isotope systematics of presolar silicate grains: New insights from Magnesium and Silicon. *Astrophys. J.* **913**, 10–27.

Huss G. R. (1990) Ubiquitous interstellar diamond and SiC in primitive chondrites: abundances reflect metamorphism. *Nature* **347**, 159–162.

Huss G. R. and Lewis R. S. (1995) Presolar diamond, SiC, and graphite in primitive chondrites: Abundances as a function of meteorite class and petrologic type. *Geochim. Cosmochim. Acta* **59**, 115–160.

Huss G. R., Meshik A. P., Smith J. B. and Hohenberg C. M. (2003) Presolar diamond, silicon carbide, and graphite in carbonaceous chondrites: Implications for thermal processing in the solar nebula. *Geochim. Cosmochim. Acta* **67**, 4823–4848.

Kruijer T. S., Kleine T. and Borg L. E. (2019) The great isotopic dichotomy of the early solar system. *Nature Astron.* **4**, 32–40.

Le Guillou C., Bernard S., Brearley A. J. and Remusat L. (2014) Evolution of organic matter in Orgueil and Renazzo during parent body aqueous alteration: In situ investigations. *Geochim. Cosmochim. Acta* **131**, 368–392.

Leitner J., Vollmer C., Hoppe P. and Zipfel J. (2012) Characterization of presolar material in the CR chondrite Northwest Africa 852. *Astrophys. J.* **745**, 38–54.

Leitner J., Metzler K. and Hoppe P. (2014) Characterization of presolar grains in cluster chondrite clasts from unequilibrated ordinary chondrites. *Lunar Planet. Sci. Conf.* **45**, abstr. #1099.

Leitner J. and Hoppe P. (2019) A new population of dust from stellar explosions among meteoritic stardust. *Nature Astron.* **3**, 725–729.

Leitner J., Metzler K., Vollmer C., Floss C., Haenecour P., Kodolányi J., Harries D. and Hoppe P. (2020) The presolar grain inventory of fine-grained chondrule rims in the Mighei-type (CM) chondrites. *Meteorit. Planet. Sci.* **55**, 1176–1206.

Liu N., Stephan T., Cristallo S., Gallino R., Boehnke P., Nittler L. R., Alexander C. M. O'D., Davis A. M., Trappitsch R., Pellin M. J. and Dillmann I. (2019) Presolar silicon carbide grains of types Y and Z: Their molybdenum isotopic compositions and stellar origins. *Astrophys. J.* **881**, 28–42.

Liu N., Barosch J., Nittler L. R., Alexander C. M. O'D., Wang J., Cristallo S., Busso M. and Palmerini S. (2021) New multielement isotopic compositions of presolar SiC grains: Implications for their stellar origins. *Astrophys. J. Lett.* **920**, L26.

Lugaro M., Zinner E., Gallino R. and Amari S. (1999) Si isotopic ratios in mainstream presolar SiC grains revisited. *Astrophys. J.* **527**, 369–394.

Matrajt G., Messenger S., Brownlee D. and Joswiak D. (2012) Diverse forms of primordial organic matter identified in interplanetary dust particles. *Meteorit. Planet. Sci.* **47**, 525–549.

Mostefaoui S., Hoppe P., Marhas K. K. and Gröner E. (2003) Search for in situ presolar oxygen-rich dust in meteorites. *66th Annual Meeting of the Meteoritical Society*, abstr. #5185.

Mostefaoui S., Marhas K. K. and Hoppe P. (2004) Discovery of an in-situ presolar silicate grain with GEMS-like composition in the Bishunpur matrix. *Lunar Planet. Sci. Conf.* **35**, abstr. #1593.

Nguyen A. N., Stadermann F. J., Zinner E., Stroud R. M., Alexander C. M. O'D. and Nittler L. R. (2007) Characterization of presolar silicate and oxide grains in primitive carbonaceous chondrites. *Astrophys. J.* **656**, 1223–1240.

Nguyen A. N., Nittler L. R., Stadermann F. J., Stroud R. M. and Alexander C. M. O'D. (2010) Coordinated analyses of presolar grains in Allan Hills 77307 and Queen Elizabeth Range 99177 meteorites. *Astrophys. J.* **719**, 166–189.

Nittler L. R., Alexander C. M. O'D., Gao X., Walker R. M. and Zinner E. K. (1994) Interstellar oxide grains from the Tieschitz ordinary chondrite. *Nature* **370**, 443–446.

Nittler L. R., Alexander C. M. O'D., Gao X., Walker R. M. and Zinner E. (1997) Stellar sapphires: The properties and origins of presolar $Al_2O_3$ in meteorites. *Astrophys. J.* **483**, 475–495.

Nittler L. R., Alexander C. M. O'D., Wang J. and Gao X. (1998) Meteoritic oxide grain from supernova found. *Nature* **393**, 222.

Nittler L. R., Alexander C. M. O'D., Gallino R., Hoppe P., Nguyen A., Stadermann F. J. and Zinner E. K. (2008) Aluminium-, calcium- and titanium-rich oxide stardust in ordinary chondrite meteorites. *Astrophys. J.* **682**, 1450–1478.

Nittler L. R., Alexander C. M. O'D., Davidson J., Riebe M. E. I., Stroud R. M. and Wang J. (2018a) High abundances of presolar grains and $^{15}$N-rich organic matter in CO3.0 chondrite Dominion Range 08006. *Geochim. Cosmochim. Acta* **226**, 107–131.

Nittler L. R., Alexander C. M. O'D, Liu N. and Wang J. (2018b) Extremely 54Cr and 50Ti-rich presolar oxide grains in a primitive meteorite: Formation in rare types of supernovae and implications for the astrophysical context of Solar System birth. *Astrophys. J. Lett.* **856**, L24.

Nittler L. R., Stroud R. M., Alexander C. M. O'D. and Howell K. (2019) Presolar grains in primitive ungrouped carbonaceous chondrite Northwest Africa 5958. *Meteorit. Planet. Sci.* **55**, 1160–1175.

Nittler L. R., Alexander C. M. O'D., Patzer A. and Verdier-Paoletti M. J. (2021) Presolar stardust in highly pristine CM chondrites Asuka 12169 and Asuka 12236. *Meteorit. Planet. Sci.* **56**, 260–276.

Ogliore R., Nagashima K., Huss G. and Haenecour P. (2021) A reassessment of the quasi-simultaneous arrival effect in secondary ion mass spectrometry. *Nuclear Instruments and Methods in Physics Research B* **491**, 17–28.

Qin L., Nittler L. R., Alexander C. M. O'D., Wang J., Stadermann F. J. and Carlson R. W. (2011) Extreme $^{54}$Cr-rich nano-oxides in the CI chondrite Orgueil – Implications for a late supernova injection into the solar system. *Geochim. Cosmochim. Acta* **75**, 629–644.

Piani L., Robert F., Beyssac O., Binet L., Bourot-Denise M., Derenne S., Le Guillou C., Marrocchi Y., Mostefaoui S., Rouzaud J.-N. and Thomen A. (2012) Structure, composition and location of the organic matter found in the enstatite chondrite Sahara 97096 (EH3). *Meteorit. Planet. Sci.* **47**, 8–29.

Piani L., Marrocchi Y., Vacer L. G., Yurimoto H. and Bizzarro M (2021) Origin of hydrogen isotopic variations in chondritic water and organics. *Earth Planet. Sci. Lett.* **567**, 117008.

Remusat L., Piani L. and Bernard S. (2016) Thermal recalcitrance of the organic D-rich component of ordinary chondrites. *Earth Planet. Sci. Lett.* **435**, 36–44.

Remusat L., Bonnet J-Y., Bernard S., Buch A. and Quirico E. (2019) Molecular and isotopic behavior of insoluble organic matter of the Orgueil meteorite upon heating. *Geochim. Cosmochim. Acta* **263**, 235–247.

Riebe M. E. I., Busemann H., Alexander C. M. O'D., Nittler L. R., Herd C. D. K., Maden C., Wang J. and Wieler R. (2020) Effects of aqueous alteration on primordial noble gases and presolar SiC in the carbonaceous chondrite Tagish Lake. *Meteorit. Planet. Sci.* **55**, 1257–1280.

Scott E. R. D. and Krot A. N. (2014) Chondrites and their components. In: *Meteorites and Cosmochemical Processes* (ed. A. M. Davis), Vol. 1, Treatise on Geochemistry 2$^{nd}$ ed. (exec. eds. H. D. Holland & K. K. Turekian), 65–137.

Singerling S. A., Nittler L. R., Barosch J., Dobrică E., Brearley A. J. and Stroud R. M. (2022) TEM analyses of in situ presolar grains from unquilibrated ordinary chondrite LL3.0 Semarkona. *Geochim. Cosmochim. Acta* **328**, 130–152.

Slodzian G., Hillion F., Stadermann F. J. and Zinner E. (2004) QSA influences on isotopic ratio measurements. *Appl. Surf. Sci.* **231**, 874–877.

Stadermann F. J., Croat T. K., Bernatowicz S., Amari S., Messenger S., Walker R. M. and Zinner E. (2005) Supernova graphite in the NanoSIMS: Carbon, oxygen and titanium isotopic compositions and its TiC sub-components. *Geochim. Cosmochim. Acta* **69**, 177–188.

Stephan T., Bose M., Boujibar A., Davis A. M., Dory C. J., Gyngard F., Hoppe P., Hynes K. M., Liu N. and Nittler L. R. (2020) The presolar grain database reloaded – Silicon Carbide. *Lunar Planet. Sci. Conf.* **51**, abstr. #2140.

Sugiura N. and Fujiya W. (2014) Correlated accretion ages and $\varepsilon^{54}$Cr of meteorite parent bodies and the evolution of the solar nebula. *Meteorit. Planet. Sci.* **49**, 772–787.

Tonotani A., Kobayashi S., Nagashima K., Sakamoto N., Russell S. S., Itoh S. and Yurimoto H. (2006) *Lunar Planet. Sci. Conf.* **37**, abstr. #1539.

Verdier-Paoletti M. J., Nittler L. R. and Wang J. (2020) New estimation of presolar grain abundances in the Paris meteorite. *Lunar Planet. Sci. Conf.* **51**, abstr. #2523.

Vollmer C., Hoppe P. and Brenker F. E. (2008) Si isotopic compositions of presolar silicate grains from red giant stars and supernovae. *Astrophys. J.* **684**, 611–617.

Vollmer C., Hoppe P., Stadermann F. J., Floss C. and Brenker F. E. (2009) NanoSIMS analysis and Auger electron spectroscopy of silicate and oxide stardust from the carbonaceous chondrite Acfer 094. *Geochim. Cosmochim. Acta* **73**, 7127–7149.

Warren P. H. (2011) Stable-isotopic anomalies and the accretionary assemblage of the Earth and Mars: A subordinate role for carbonaceous chondrites. *Earth Planet. Sci. Lett.* **311**, 93–100.

Wasson J. T. and Rubin A. E. (2009) Composition of matrix in the CR chondrite LAP 02342. *Geochim. Cosmochim. Acta* **73**, 1436–1460.

Yada T., Floss C., Stadermann F. J., Zinner E., Nakamura T., Noguchi T. and Lea A. S. (2008) Stardust in Antarctic micrometeorites. *Meteorit. Planet. Sci.* **43**, 1287–1298.

Zhao X., Floss C., Lin Y. and Bose M. (2013) Stardust investigation into the CR chondrite Grove Mountain 021710. *Astrophys. J.* **769**, 49–65.

Zhao X., Lin Y., Yin Q.-Z., Zhang J., Hao J., Zolensky M. and Jenniskens P. (2014) Presolar grains in the CM2 chondrite Sutter's Mill. *Meteorit. Planet. Sci.* **49**, 2038–2046.

Zinner E., Nittler L. R., Hoppe P., Gallino R., Straniero O. and Alexander C. M. O'D. (2005) Oxygen, magnesium and chromium isotopic ratios of presolar spinel grains. *Geochim. Cosmochim. Acta* **69**, 4149–4165.

Zinner E. (2014) Presolar grains. In: *Meteorites and Cosmochemical Processes* (ed. A. M. Davis), Vol. 1, Treatise on Geochemistry 2nd ed. (exec. eds. H. D. Holland & K. K. Turekian), 181–213.

# Tables

Table 1: UOC matrix areas analyzed with the NanoSIMS and matrix-normalized abundances of presolar O-anomalous grains, SiC grains and carbonaceous grains.

| | Targeted | Measured area | **O-anom. grains** | | | **SiC grains** | | | **Carbonaceous grains** | | |
|---|---|---|---|---|---|---|---|---|---|---|---|
| Area | area* | ($\mu m^2$)** | #grains | abun. (ppm) | error +/- | #grains | abun. (ppm) | error +/- | #grains | abun. (ppm) | error +/- |
| Sem F1 | TEM C90A | 3000 | 3 | 152.1 | 148/83 | 2 | 38.7 | 51/25 | 2 | 185.1 | 244/120 |
| Sem F2A1 | TEM C90B | 11187 | 15 | 107.4 | 42/33$^{MC}$ | 8 | 53.7 | 27/19 | 10 | 51.7 | 22/16 |
| Sem F2A2 | TEM C89 | 12133 | 15 | 191.5 | 98/89$^{MC}$ | 11 | 56.2 | 23/17 | 3 | 27.9 | 27/15 |
| **Semarkona** | **Total:** | **26320** | **33** | **151.3** | **50/46$^{MC}$** | **21** | **53.1** | **14/12** | **15** | **56.0** | **19/14** |
| | | | | | | | | | | | |
| MET A01 | SEM | 5264 | 4 | 52.9 | 42/25 | 1 | 23.5 | 54/19 | 0 | - | - |
| MET A02 | SEM | 9910 | 4 | 34.3 | 27/16 | 4 | 27.5 | 22/13 | 1 | 6.5 | 15/5 |
| MET A05 | SEM | 5146 | 9 | 104.6 | 48/34 | 1 | 5.2 | 12/4 | 0 | - | - |
| MET A07 | SEM | 7975 | 6 | 39.8 | 24/16 | 8 | 49.5 | 24/17 | 2 | 6.5 | 9/4 |
| MET A08 | SEM | 5282 | 3 | 54.6 | 53/30 | 1 | 11.8 | 27/10 | 0 | - | - |
| MET A09 | SEM | 5594 | 8 | 75.9 | 37/26 | 4 | 31.1 | 25/15 | 0 | - | - |
| MET A10 | SEM | 7191 | 7 | 34.4 | 19/13 | 0 | - | - | 0 | - | - |
| MET A11 | SEM | 5607 | 3 | 11.4 | 11/6 | 1 | 11.2 | 26/9 | 1 | 12.5 | 29/10 |
| MET A12 | SEM | 3427 | 10 | 148.9 | 64/46 | 2 | 35.0 | 46/23 | 0 | - | - |
| MET A13 | SEM | 2851 | 3 | 72.6 | 71/40 | 1 | 20.9 | 48/17 | 0 | - | - |
| MET A14 | SEM | 2550 | 3 | 18.4 | 18/10 | 1 | 9.3 | 21/8 | 0 | - | - |
| MET-B02 | TEM C30 | 5961 | 4 | 22.0 | 17/11 | 1 | 5.6 | 13/5 | 0 | - | - |
| MET-B03 | TEM C28 | 4155 | 5 | 164.8 | 111/71 | 0 | 0.0 | - | 0 | - | - |
| MET-B04 | SEM | 3950 | 3 | 35.8 | 35/19 | 5 | 26.4 | 18/11 | 1 | 8.0 | 18/7 |
| MET-B06 | SEM | 2850 | 4 | 74.6 | 59/36 | 3 | 42.2 | 41/23 | 0 | - | - |
| MET-C01 | SEM | 1419 | 1 | 87.1 | 200/72 | 2 | 23.8 | 31/15 | 0 | - | - |
| MET-C02 | SEM | 2335 | 0 | - | - | 3 | 51.1 | 50/28 | 0 | - | - |
| MET-C03 | SEM | 2398 | 1 | 7.4 | 17/6 | 3 | 31.0 | 30/17 | 0 | - | - |
| **MET 00526** | **Total:** | **83865** | **78** | **54.5** | **11/10$^{MC}$** | **41** | **21.5** | **5/4$^{MC}$** | **5** | **2.6** | **2/1** |
| | | | | | | | | | | | |
| NWA A01 | SEM | 1523 | 1 | 41.1 | 95/34 | 1 | 23.6 | 54/20 | 0 | - | - |
| NWA A04 | SEM | 2567 | 0 | - | - | 1 | 19.6 | 45/16 | 0 | - | - |
| NWA A05 | SEM | 2773 | 0 | - | - | 1 | 20.9 | 48/17 | 0 | - | - |
| NWA A07 | SEM | 2625 | 2 | 24.8 | 33/16 | 0 | - | - | 0 | - | - |
| NWA A08 | SEM | 3207 | 3 | 48.6 | 67/26 | 1 | 7.4 | 17/6 | 0 | - | - |
| NWA A11 | SEM | 3229 | 0 | - | - | 1 | 16.1 | 37/13 | 1 | 10.2 | 23/8 |
| NWA A13 | SEM | 4574 | 0 | - | - | 2 | 25.7 | 34/17 | 0 | - | - |
| NWA A14 | SEM | 3771 | 1 | 8.7 | 20/7 | 0 | - | - | 0 | - | - |
| NWA B03 | SEM | 4029 | 1 | 7.4 | 17/6 | 0 | - | - | 0 | - | - |
| NWA B04 | SEM | 1064 | 0 | - | - | 2 | 105.5 | 139/68 | 0 | - | - |
| **NWA 8276** | **Total:** | **29363** | **8** | **11.8** | **6/3$^{MC}$** | **9** | **15.3** | **7/5** | **1** | **1.1** | **3/1** |

*Targeted area: TEM = these NanoSIMS maps were placed directly adjacent to areas studied with TEM (Dobrică and Brearley 2020a, b). The numbers indicate FIB section nomenclature: for C89, C90A, C90B see Figs. 2–6 in Dobrică and Brearley (2020a); for C28 and C30 see Figs. 1, 2, 5–7 in Dobrică and Brearley (2020b). SEM = areas were chosen with the SEM, avoiding clast-rich and/or visibly altered matrix.

**Measured area after corrections (cf. Section 2).

Uncertainties are 1σ and were determined from the confidence limits for small numbers of events provided by Gehrels (1986) or, if indicated ($^{MC}$), by means of the Monte Carlo method of Nittler et al. (2018a). Abun. = abundances.

**Figures**

Fig. 1:

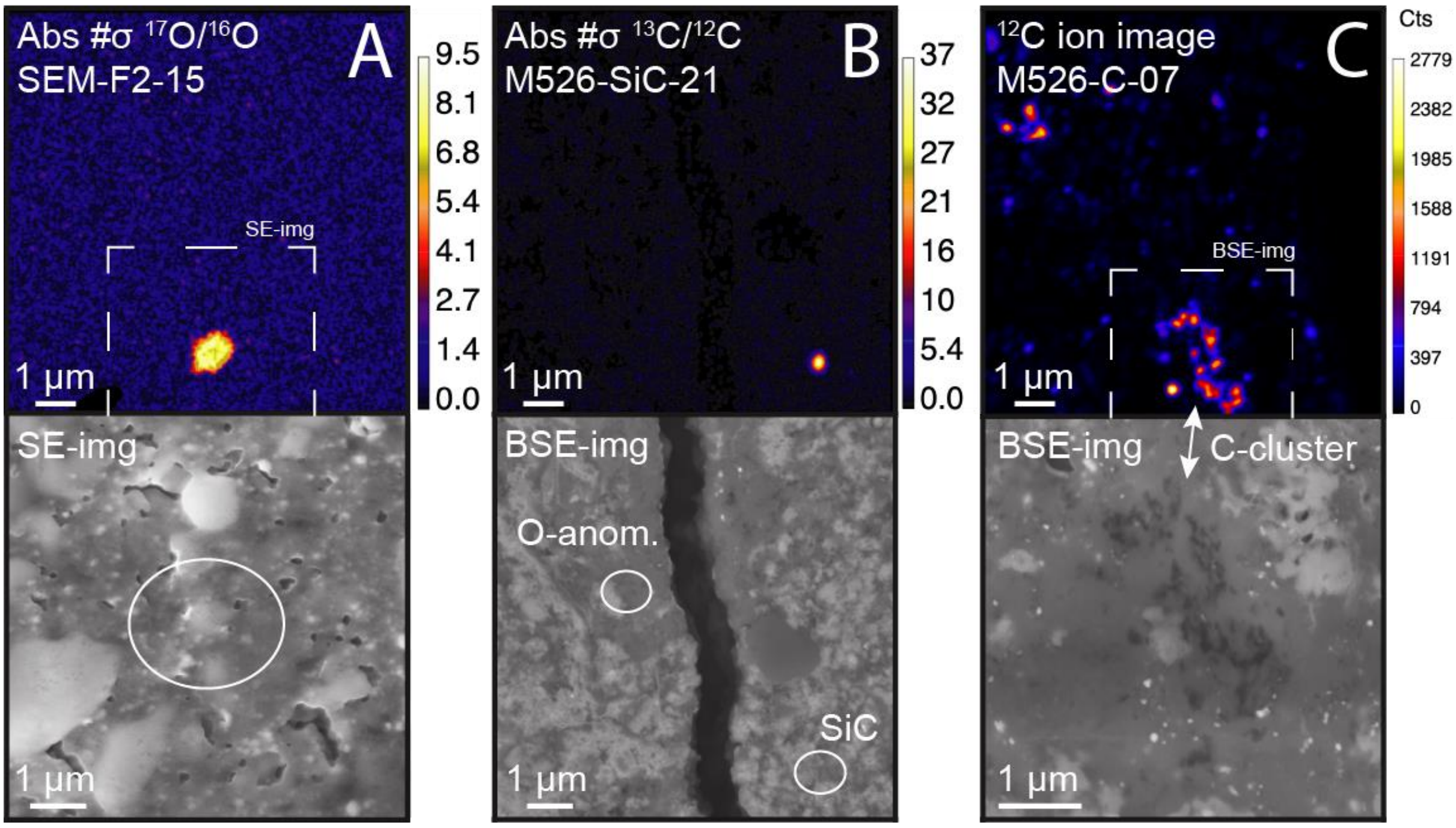

Gallery of NanoSIMS and SEM images of presolar grains. Image A shows a large (0.9 µm diameter) $^{17}$O-rich presolar grain and image B shows a SiC grain with $\delta^{13}$C>26,000‰. Image C is a cluster consisting of at least nine C-rich and partly C-anomalous organic grains.
SE = secondary electron image, BSE = backscattered electron image, Abs #σ = 'sigma image', shows for every pixel in the image the number of σ deviation from the matrix isotope ratio.

Fig. 2:

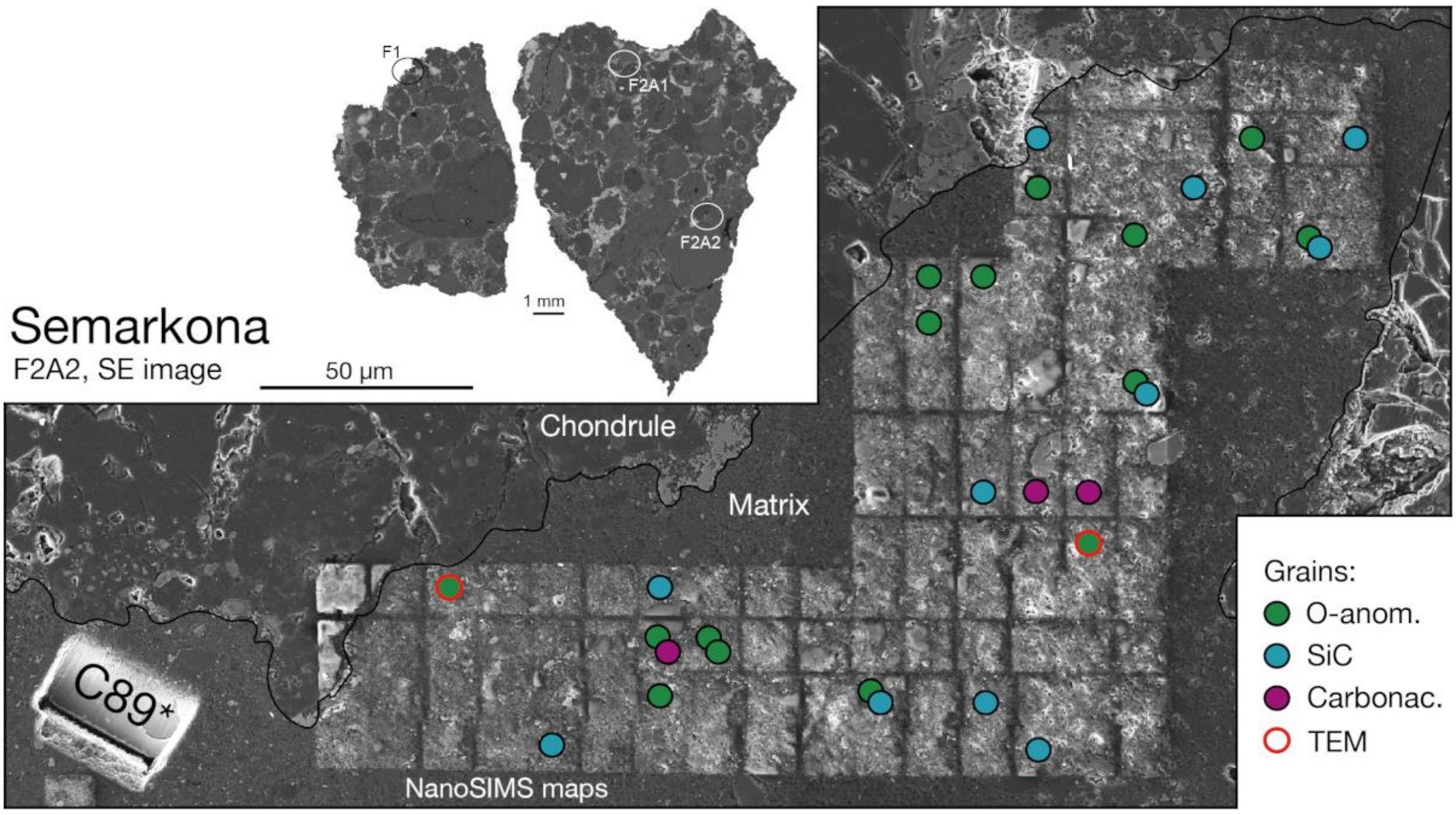

Secondary electron (SE) images of Semarkona thin sections and matrix area F2A2. The NanoSIMS maps were placed directly next to a focused ion beam section studied with TEM that revealed abundant amorphous silicates in this area (Dobrică and Brearley 2020a). The high abundance of presolar O-anomalous grains (~192 ppm) confirms that this matrix is relatively pristine. The two grains outlined in red, as well as five more grains from other Semarkona areas, were examined with TEM (Singerling et al. 2022).

*C89 = focused ion beam section studied with TEM (see Figs. 4 and 5a in Dobrică and Brearley 2020a).

Fig. 3:

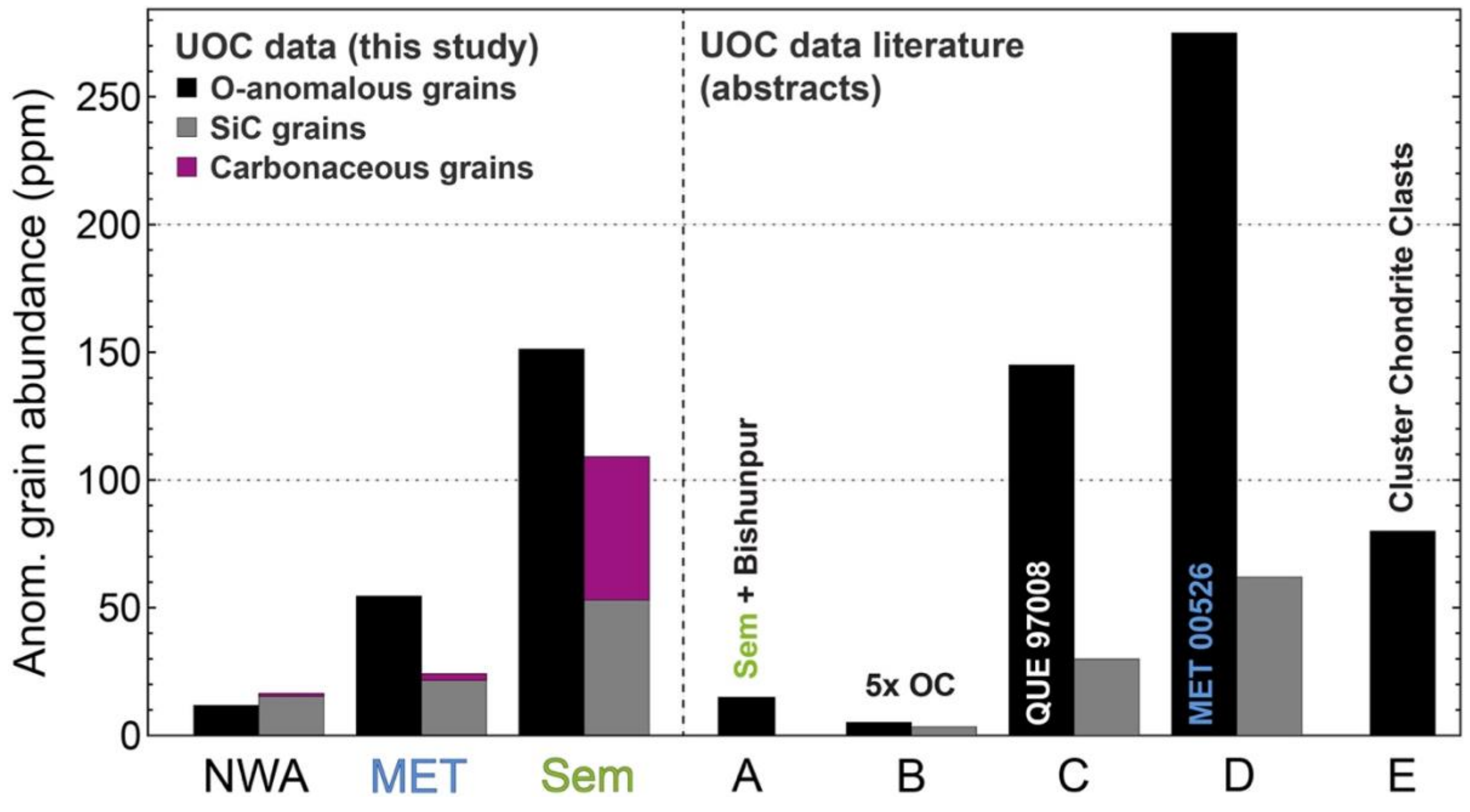

Average O- and C-anomalous grain abundances in Semarkona, MET 00526, NWA 8276 and other UOCs. Literature data was taken from the following abstracts: A = Mostefaoui et al. (2003, 2004); B = Tonotani et al. (2006), average presolar grain abundance of five UOCs: Semarkona, Bishunpur, Krymka, Jiddat al Harasis 026 and Dhofar 008; C = Floss and Haenecour (2016c) D = Floss and Haenecour (2016b); E = Leitner et al. (2014), average presolar grains abundance in fine-grained chondrule rims of cluster chondrite clasts in NWA 1756 and Krymka. Anom. = anomalous.

Fig. 4:

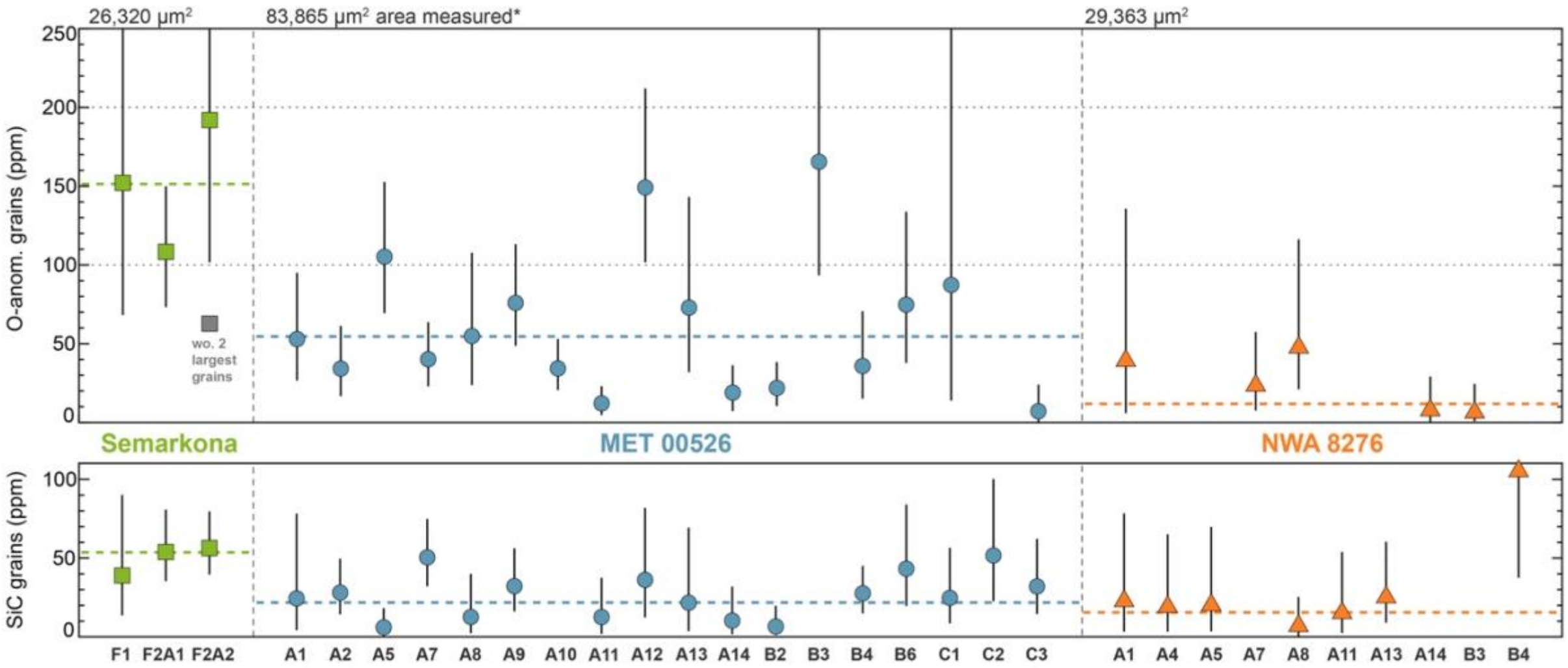

The matrix-normalized abundances of presolar O-anomalous and SiC grains in individual Semarkona, MET 00526 and NWA 8276 areas. Areas without O-anomalous or SiC grains are not displayed (cf. Table 1). All Semarkona maps and MET maps B02 and B03 were placed adjacent to areas studied with TEM (Dobrică and Brearley 2020a, b). The other MET areas and all NWA areas were chosen with the SEM and may be less pristine. Uncertainties are 1σ (Gehrels 1986; Nittler et al. 2018a). The colored horizontal lines represent average presolar grain abundances in the UOCs studied. Anom. = anomalous; wo. = without.

*Measured area after corrections (cf. Section 2).

Fig. 5:

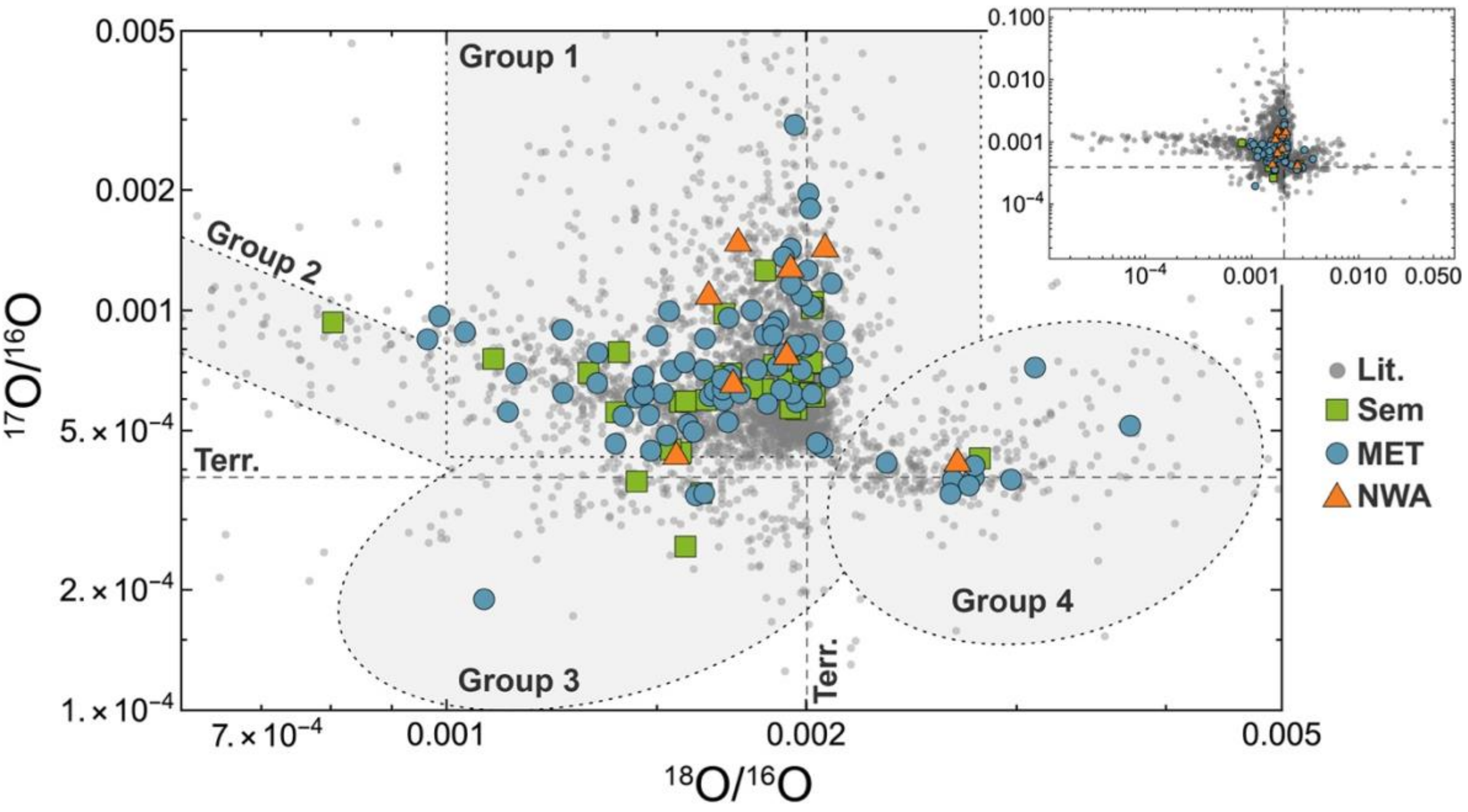

Oxygen three-isotope plot of O-anomalous presolar grains in unequilibrated ordinary chondrites Semarkona, MET 00526, and NWA 8276. Literature data of presolar oxides and silicates in mainly CC and C-ungrouped chondrites are from a large number of sources (e.g., Nittler et al. 1997, 2008, 2018a, 2021; Vollmer et al. 2009; Nguyen et al. 2007, 2010; Leitner et al. 2012; Zhao et al. 2013; Zinner 2014; Floss and Haenecour 2016a; Haenecour et al. 2018; Hoppe et al. 2021, and references therein). The inlet shows the full range of compositions reported in the literature. Note that the most anomalous compositions were measured in oxides from acid residues, whereas in situ NanoSIMS studies only provide a lower limit of the anomalies.

Fig. 6:

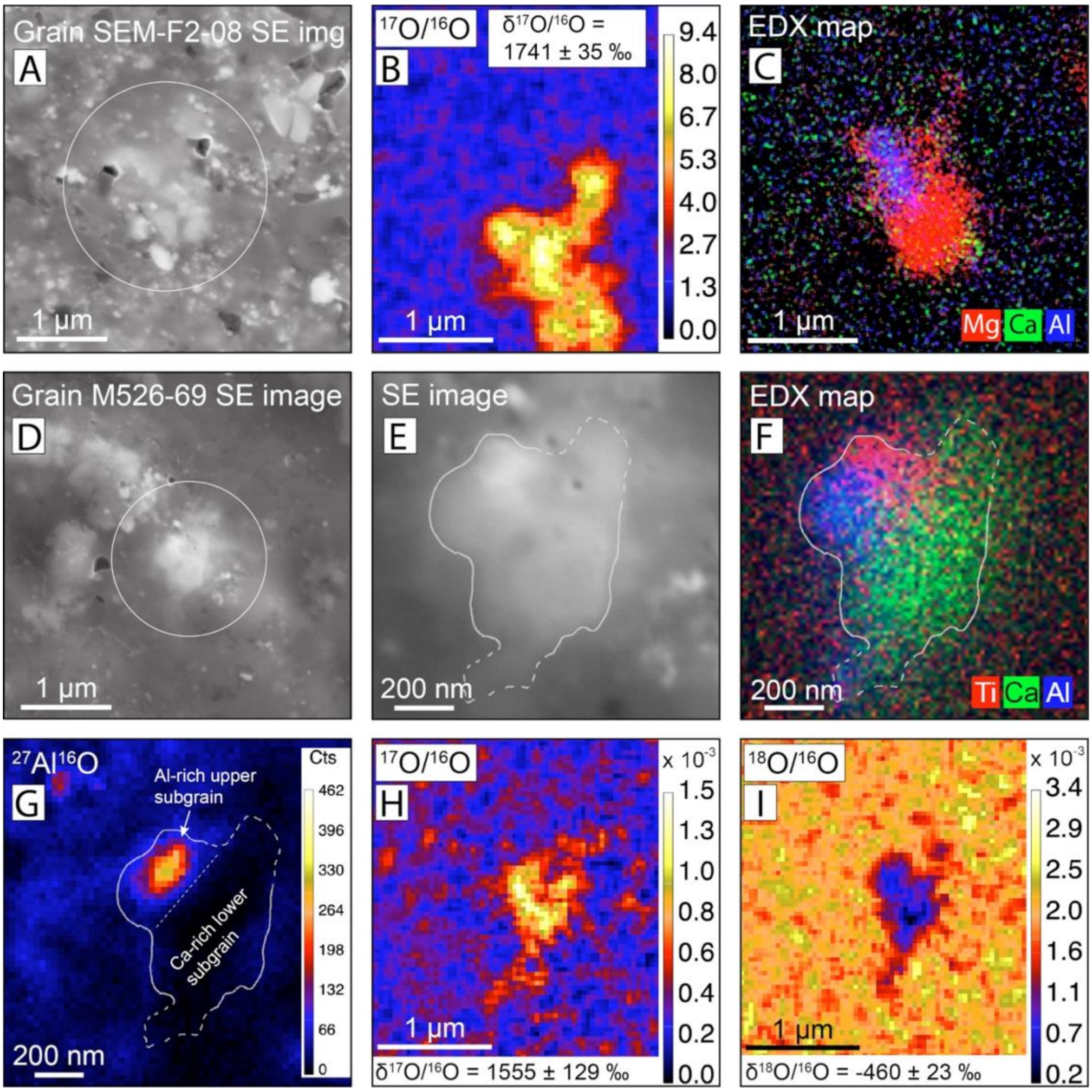

This gallery shows two composite presolar grains made from oxide and silicate phases. The first row (images A–C) shows an SE image, a $^{17}O/^{16}O$ ratio image and an EDX map of the amoeboidal-shaped composite grain SEM-F2-08. Images D–I show grain M526-69. The EDX composite map and $^{27}Al^{16}O$ ion image reveal a Ti-Al-rich upper oxide subgrain and a Ca-rich lower silicate subgrain.

Fig. 7:

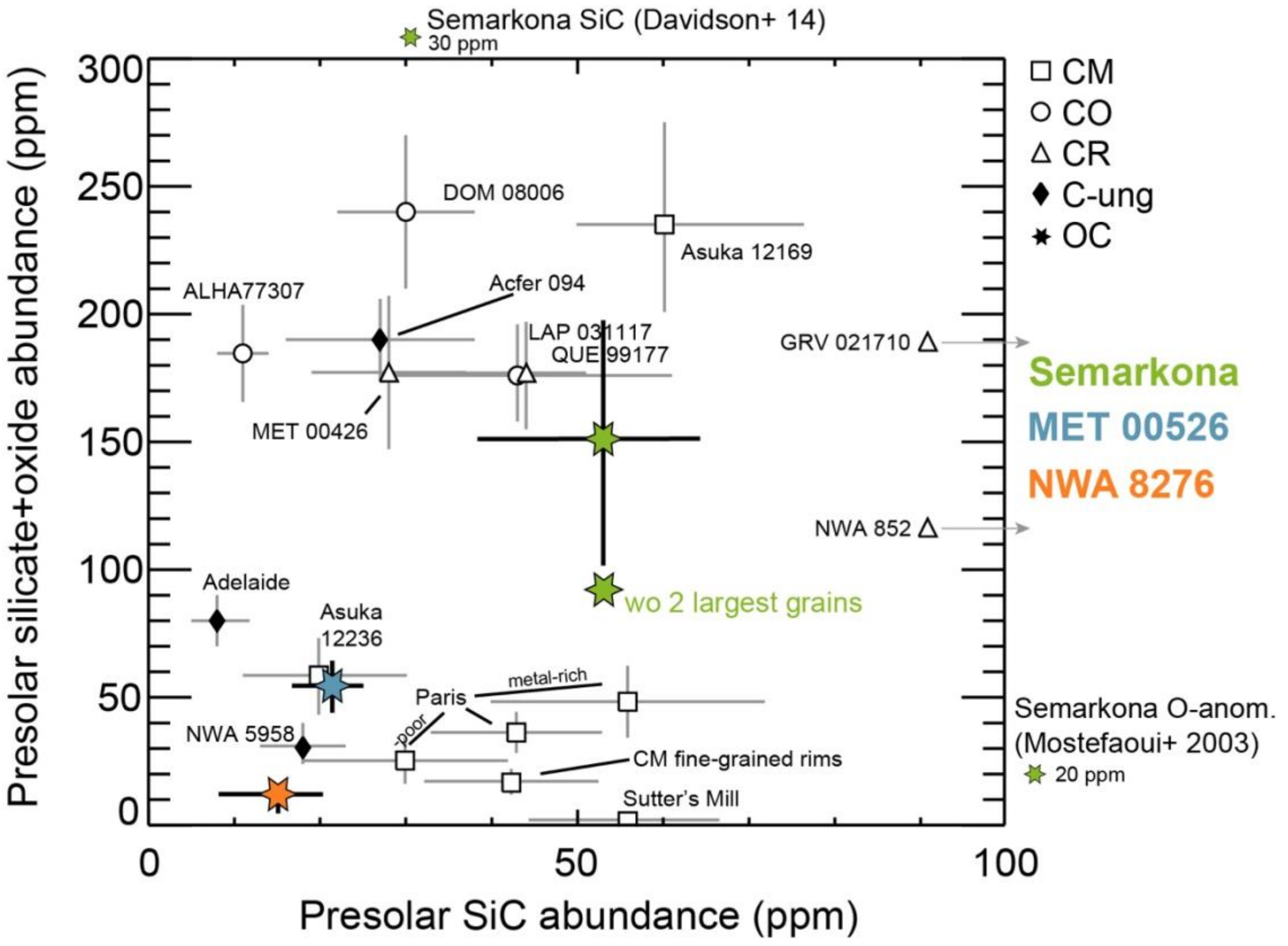

The matrix-normalized abundances of presolar O-anomalous and SiC grains in Semarkona, MET 00526, and NWA 8276 compared to those in CC and C-ungrouped chondrites. ALH = Allan Hills; GRA = Grave Nunataks; GRV = Grove Mountain; LAP = La Paz Icefield; NWA = Northwest Africa; QUE = Queen Alexandra Range. Data from Mostefaoui et al. (2003); Floss and Stadermann (2009b, 2012); Nguyen et al. (2010); Leitner et al. (2012, 2020); Zhao et al. (2013, 2014); Davidson et al. (2014); Haenecour et al. (2018); Nittler et al. (2018a, 2019, 2021); Riebe et al. (2020); Verdier-Paoletti et al. (2020). Uncertainties are 1σ. Wo. = without.

# Supplementary Material

# Presolar O- and C-anomalous grains in unequilibrated ordinary chondrite matrices

Jens Barosch[1*], Larry R. Nittler[1], Jianhua Wang[1], Elena Dobrică[2], Adrian J. Brearley[3], Dominik C. Hezel[4], Conel M. O'D. Alexander[1]

[1]Earth and Planets Laboratory, Carnegie Institution of Washington, 5241 Broad Branch Rd. NW, Washington DC 20015, USA.

[2]Hawai'i Institute of Geophysics and Planetology, University of Hawai'i at Mānoa, 1680 East-West Road, Honolulu 96822, USA.

[3]Department of Earth and Planetary Sciences, 1 University of New Mexico, Albuquerque, New Mexico 87131-0001, USA.

[4]Institut für Geowissenschaften, Goethe-Universität Frankfurt, Altenhöferallee 1, 60438 Frankfurt am Main, Germany.

**corresponding author:* jbarosch@carnegiescience.edu

Link to article

## Contents:

A Table listing the characteristics of all presolar O- and C-anomalous grains identified in this study (Table S1).

NanoSIMS image simulations are presented to examine the effects of primary ion beam tailing and counting statistics on the identification and quantification of presolar grains.

**Table S1:** A - Characteristics of presolar O-anomalous grains identified in Semarkona, MET 00526 and NWA 8276.

| Grain | Region | Size (µm)* | Size corr | Group | $^{17}O/^{16}O$ | $^{17}O/^{16}O$ error | $^{18}O/^{16}O$ | $^{18}O/^{16}O$ error | $^{28}Si^-/^{16}O^-$ | $^{27}Al^{16}O^-/^{16}O^-$ | Inferred Phase | Comments |
|---|---|---|---|---|---|---|---|---|---|---|---|---|
| SEM-F2-02 | F2A2 | 0.31 | 0.31 | 4 | 0.000426 | 0.000033 | 0.002795 | 0.000083 | 0.025498 | 0.003708 | Silicate | |
| SEM-F2-03 | F2A2 | 0.13 | 0.06 | 1 | 0.000786 | 0.000087 | 0.001394 | 0.000138 | 0.023974 | 0.010729 | Silicate | |
| SEM-F2-04 | F2A2 | 0.25 | 0.22 | 1 | 0.000609 | 0.000042 | 0.002033 | 0.000076 | 0.026258 | 0.003309 | Silicate | |
| SEM-F2-05 | F2A2 | 0.28 | 0.28 | 1 | 0.000635 | 0.000037 | 0.001842 | 0.000065 | 0.023829 | 0.005546 | Silicate | |
| SEM-F2-06 | F2A2 | 0.25 | 0.22 | 1 | 0.000441 | 0.000035 | 0.001572 | 0.000073 | 0.023794 | 0.005590 | Silicate | |
| SEM-F2-08 | F2A2 | 1.09 | 1.09 | 1 | 0.001049 | 0.000013 | 0.002028 | 0.000018 | 0.023608 | 0.005190 | Composite | TEM, F2-8 |
| SEM-F2-09 | F2A2 | 0.41 | 0.41 | 2 | 0.000934 | 0.000033 | 0.000804 | 0.000047 | 0.029428 | 0.002989 | Silicate | TEM, F2-9 in Singerling et al. (2022) |
| SEM-F2-11 | F2A2 | 0.37 | 0.37 | 1 | 0.000698 | 0.000029 | 0.001315 | 0.000049 | 0.031536 | 0.004098 | Silicate | |
| SEM-F2-12 | F2A2 | 0.29 | 0.29 | 1 | 0.000556 | 0.000033 | 0.001387 | 0.000062 | 0.027859 | 0.004951 | Silicate | |
| SEM-F2-13 | F2A2 | 0.34 | 0.34 | 1 | 0.000756 | 0.000035 | 0.001095 | 0.000056 | 0.029756 | 0.002442 | Silicate | |
| SEM-F2-14 | F2A2 | 0.25 | 0.22 | 1 | 0.000731 | 0.000050 | 0.001881 | 0.000081 | 0.028353 | 0.004075 | Silicate | |
| SEM-F2-15 | F2A2 | 0.90 | 0.90 | 1 | 0.001258 | 0.000018 | 0.001849 | 0.000023 | 0.027920 | 0.001615 | Silicate | |
| SEM-F2-16 | F2A2 | 0.22 | 0.18 | 1 | 0.000598 | 0.000041 | 0.001650 | 0.000074 | 0.033019 | 0.008457 | Silicate | |
| SEM-F2-17 | F2A2 | 0.22 | 0.19 | 1 | 0.000615 | 0.000044 | 0.001955 | 0.000077 | 0.025072 | 0.009507 | Silicate | |
| SEM-F2-18 | F2A2 | 0.27 | 0.27 | 1 | 0.000736 | 0.000044 | 0.002009 | 0.000072 | 0.031434 | 0.007557 | Silicate | |
| SEM-F2-20 | F2A1 | 0.38 | 0.38 | 1 | 0.000449 | 0.000021 | 0.001541 | 0.000044 | 0.027133 | 0.001941 | Silicate | |
| SEM-F2-22 | F2A1 | 0.32 | 0.32 | 1 | 0.000755 | 0.000036 | 0.001925 | 0.000058 | 0.021743 | 0.001123 | Silicate | |
| SEM-F2-23 | F2A1 | 0.37 | 0.37 | 1 | 0.000640 | 0.000024 | 0.001802 | 0.000042 | 0.029816 | 0.003302 | Silicate | TEM, F2-23 in Singerling et al. (2022) |
| SEM-F2-24 | F2A1 | 0.24 | 0.21 | 1 | 0.000695 | 0.000041 | 0.001731 | 0.000068 | 0.027376 | 0.003205 | Silicate | |
| SEM-F2-26 | F2A1 | 0.22 | 0.18 | 3 | 0.000373 | 0.000047 | 0.001443 | 0.000106 | 0.027839 | 0.000047 | Silicate | |
| SEM-F2-28 | F2A1 | 0.25 | 0.22 | 1 | 0.000565 | 0.000033 | 0.001963 | 0.000061 | 0.019243 | 0.002007 | Silicate | |
| SEM-F2-29 | F2A1 | 0.66 | 0.66 | 1 | 0.000981 | 0.000016 | 0.001709 | 0.000023 | 0.026318 | 0.002427 | Silicate | |
| SEM-F2-30 | F2A1 | 0.32 | 0.32 | 1 | 0.001014 | 0.000036 | 0.002022 | 0.000050 | 0.025186 | 0.001521 | Silicate | TEM, F2-30b in Singerling et al. (2022) |
| SEM-F2-31 | F2A1 | 0.28 | 0.28 | 1 | 0.000620 | 0.000040 | 0.002016 | 0.000071 | 0.012905 | 0.002952 | Silicate | |
| SEM-F2-32 | F2A1 | 0.32 | 0.32 | 1 | 0.000569 | 0.000030 | 0.001940 | 0.000055 | 0.021194 | 0.001744 | Silicate | |
| SEM-F2-33 | F2A1 | 0.22 | 0.18 | 3 | 0.000257 | 0.000047 | 0.001585 | 0.000106 | 0.010820 | 0.000208 | Silicate | |
| SEM-F2-34 | F2A1 | 0.26 | 0.26 | 3 | 0.000349 | 0.000026 | 0.001635 | 0.000059 | 0.029985 | 0.002061 | Silicate | |
| SEM-F2-35 | F2A1 | 0.24 | 0.20 | 1 | 0.000662 | 0.000040 | 0.001934 | 0.000068 | 0.020920 | 0.002850 | Silicate | |
| SEM-F2-36 | F2A1 | 0.25 | 0.21 | 1 | 0.000704 | 0.000035 | 0.001989 | 0.000059 | 0.020144 | 0.001121 | Silicate | |
| SEM-F2-37 | F2A1 | 0.35 | 0.35 | 1 | 0.000741 | 0.000029 | 0.002022 | 0.000047 | 0.012818 | 0.017110 | Spinel | TEM, F2-37 in Singerling et al. (2022) |
| SEM-F1-01 | F1 | 0.65 | 0.65 | 1 | 0.000590 | 0.000013 | 0.001569 | 0.000024 | 0.026367 | 0.005402 | Silicate | TEM, F1-1 in Singerling et al. (2022) |
| SEM-F1-02 | F1 | 0.28 | 0.28 | 1 | 0.000684 | 0.000035 | 0.001678 | 0.000059 | 0.023772 | 0.001089 | Silicate | |
| SEM-F1-03 | F1 | 0.29 | 0.29 | 1 | 0.000592 | 0.000030 | 0.001586 | 0.000053 | 0.027389 | 0.002716 | Silicate | |
| M526-01 | AX | 0.42 | 0.42 | 4 | 0.000416 | 0.000016 | 0.002336 | 0.000037 | 0.024596 | 0.003689 | Silicate | out of focus, excluded from abundance calculation |
| M526-02 | AX | 0.39 | 0.39 | 1 | 0.000454 | 0.000015 | 0.002065 | 0.000032 | 0.018818 | 0.008129 | Silicate | out of focus, excluded from abundance calculation |
| M526-03 | AX | 0.43 | 0.43 | 1 | 0.000782 | 0.000017 | 0.001338 | 0.000027 | 0.017538 | 0.005014 | Silicate | out of focus, excluded from abundance calculation |
| M526-04 | A02 | 0.45 | 0.45 | 1 | 0.000466 | 0.000013 | 0.002042 | 0.000027 | 0.011934 | 0.006703 | Silicate | |
| M526-05 | A02 | 0.26 | 0.26 | 1 | 0.000620 | 0.000032 | 0.001519 | 0.000056 | 0.018566 | 0.001765 | Silicate | |
| M526-06 | A02 | 0.31 | 0.31 | 1 | 0.000519 | 0.000022 | 0.001593 | 0.000042 | 0.019328 | 0.006188 | Silicate | |
| M526-07 | A02 | 0.25 | 0.25 | 1 | 0.001960 | 0.000070 | 0.002009 | 0.000069 | 0.012832 | 0.003714 | Silicate | |
| M526-08 | A05 | 0.42 | 0.42 | 1 | 0.000589 | 0.000024 | 0.001963 | 0.000044 | 0.012229 | 0.011305 | Silicate | |
| M526-09 | A05 | 0.22 | 0.18 | 1 | 0.000611 | 0.000050 | 0.001661 | 0.000087 | 0.014005 | 0.004836 | Silicate | |
| M526-10 | A05 | 0.27 | 0.27 | 1 | 0.000869 | 0.000045 | 0.001845 | 0.000067 | 0.012010 | 0.003325 | Silicate | |
| M526-11 | A05 | 0.16 | 0.10 | 1 | 0.000690 | 0.000065 | 0.001730 | 0.000109 | 0.010634 | 0.005994 | Silicate | |
| M526-12 | A05 | 0.28 | 0.28 | 1 | 0.000630 | 0.000041 | 0.001676 | 0.000072 | 0.013600 | 0.007629 | Silicate | |
| M526-13 | A05 | 0.26 | 0.26 | 1 | 0.000596 | 0.000039 | 0.001705 | 0.000071 | 0.011795 | 0.006853 | Silicate | |
| M526-14 | A05 | 0.36 | 0.36 | 1 | 0.000726 | 0.000035 | 0.001893 | 0.000056 | 0.018196 | 0.003496 | Silicate | |
| M526-15 | A05 | 0.27 | 0.27 | 1 | 0.000939 | 0.000048 | 0.001895 | 0.000069 | 0.013322 | 0.007077 | Silicate | |
| M526-16 | A05 | 0.24 | 0.21 | 1 | 0.000959 | 0.000049 | 0.001723 | 0.000068 | 0.012510 | 0.001252 | Silicate | |
| M526-17 | A07 | 0.24 | 0.20 | 1 | 0.000634 | 0.000054 | 0.001703 | 0.000093 | 0.019512 | 0.001508 | Silicate | |
| M526-18 | A07 | 0.26 | 0.26 | 1 | 0.002910 | 0.000086 | 0.001956 | 0.000069 | 0.016272 | 0.005425 | Silicate | |
| M526-19 | A07 | 0.29 | 0.29 | 1 | 0.001426 | 0.000056 | 0.001942 | 0.000064 | 0.017578 | 0.004908 | Silicate | |
| M526-20 | A07 | 0.25 | 0.22 | 1 | 0.000525 | 0.000035 | 0.001720 | 0.000067 | 0.014092 | 0.003288 | Silicate | |
| M526-21 | A07 | 0.28 | 0.28 | 1 | 0.000497 | 0.000034 | 0.001609 | 0.000066 | 0.020329 | 0.005214 | Silicate | |
| M526-22 | A07 | 0.29 | 0.29 | 1 | 0.000894 | 0.000047 | 0.001250 | 0.000068 | 0.018123 | 0.002115 | Silicate | |
| M526-23 | A08 | 0.37 | 0.37 | 1 | 0.001259 | 0.000051 | 0.002006 | 0.000062 | 0.016610 | 0.002533 | Silicate | |
| M526-24 | A08 | 0.35 | 0.44 | 1 | 0.000995 | 0.000043 | 0.001536 | 0.000059 | 0.016942 | 0.011300 | Silicate | |
| M526-25 | A08 | 0.23 | 0.20 | 1 | 0.000850 | 0.000058 | 0.001646 | 0.000085 | 0.020142 | 0.006468 | Silicate | |
| M526-26 | A10 | 0.22 | 0.19 | 1 | 0.000668 | 0.000056 | 0.001461 | 0.000122 | 0.027023 | 0.012671 | Silicate | |
| M526-27 | A10 | 0.25 | 0.25 | 4 | 0.000720 | 0.000073 | 0.003109 | 0.000139 | 0.022089 | 0.002500 | Silicate | |
| M526-28 | A10 | 0.29 | 0.29 | 1 | 0.001025 | 0.000062 | 0.002021 | 0.000087 | 0.024342 | 0.003758 | Silicate | |
| M526-29 | A10 | 0.18 | 0.14 | 1 | 0.000709 | 0.000071 | 0.001643 | 0.000119 | 0.023383 | 0.003000 | Silicate | |
| M526-30 | A10 | 0.18 | 0.14 | 1 | 0.000673 | 0.000066 | 0.001701 | 0.000113 | 0.017095 | 0.003806 | Silicate | |
| M526-31 | A10 | 0.27 | 0.27 | 3 | 0.000189 | 0.000035 | 0.001075 | 0.000109 | 0.012440 | 0.004308 | Silicate | |
| M526-32 | A10 | 0.19 | 0.15 | 4 | 0.000383 | 0.000048 | 0.002762 | 0.000128 | 0.015764 | 0.010791 | Silicate | |
| M526-33 | A11 | 0.21 | 0.17 | 4 | 0.000377 | 0.000041 | 0.002654 | 0.000107 | 0.015410 | 0.008427 | Silicate | |
| M526-34 | A11 | 0.16 | 0.11 | 4 | 0.000409 | 0.000057 | 0.002766 | 0.000149 | 0.016539 | 0.007139 | Silicate | |
| M526-35 | A11 | 0.23 | 0.20 | 1 | 0.001090 | 0.000060 | 0.001983 | 0.000081 | 0.016284 | 0.040398 | Oxide | |
| M526-36 | A12 | 0.45 | 0.45 | 1 | 0.000710 | 0.000027 | 0.001983 | 0.000045 | 0.023291 | 0.009935 | Silicate | EDX |
| M526-37 | A12 | 0.21 | 0.17 | 1 | 0.000487 | 0.000046 | 0.001529 | 0.000093 | 0.022703 | 0.004272 | Silicate | |
| M526-38 | A12 | 0.21 | 0.17 | 1 | 0.000464 | 0.000047 | 0.001386 | 0.000098 | 0.025253 | 0.005107 | Silicate | |
| M526-39 | A12 | 0.24 | 0.20 | 1 | 0.000906 | 0.000056 | 0.001877 | 0.000083 | 0.020603 | 0.006027 | Silicate | |
| M526-40 | A12 | 0.22 | 0.18 | 4 | 0.000378 | 0.000037 | 0.002969 | 0.000103 | 0.021719 | 0.004108 | Silicate | |
| M526-41 | A12 | 0.25 | 0.25 | 1 | 0.000619 | 0.000043 | 0.001948 | 0.000078 | 0.026938 | 0.006821 | Silicate | |
| M526-42 | A12 | 0.29 | 0.29 | 1 | 0.000809 | 0.000046 | 0.001996 | 0.000073 | 0.027073 | 0.004924 | Silicate | |
| M526-43 | A12 | 0.24 | 0.20 | 1 | 0.000706 | 0.000059 | 0.001542 | 0.000086 | 0.030735 | 0.005477 | Silicate | |
| M526-44 | A12 | 0.29 | 0.29 | 1 | 0.000446 | 0.000032 | 0.001483 | 0.000068 | 0.025136 | 0.003039 | Silicate | |
| M526-45 | A12 | 0.23 | 0.20 | 1 | 0.000621 | 0.000047 | 0.001762 | 0.000085 | 0.017702 | 0.006956 | Silicate | |
| M526-46 | A13 | 0.23 | 0.20 | 1 | 0.001796 | 0.000089 | 0.002015 | 0.000093 | 0.017225 | 0.005195 | Silicate | |
| M526-47 | A13 | 0.38 | 0.38 | 1 | 0.000864 | 0.000036 | 0.001501 | 0.000065 | 0.020805 | 0.011781 | Silicate | EDX |
| M526-48 | A13 | 0.28 | 0.28 | 1 | 0.000658 | 0.000043 | 0.001337 | 0.000084 | 0.019216 | 0.009281 | Silicate | |
| M526-49 | A14 | 0.21 | 0.17 | 1 | 0.001169 | 0.000081 | 0.002101 | 0.000108 | 0.019214 | 0.004082 | Silicate | |
| M526-50 | A14 | 0.16 | 0.10 | 1 | 0.000888 | 0.000086 | 0.002109 | 0.000132 | 0.019936 | 0.012189 | Silicate | |
| M526-51 | A14 | 0.19 | 0.14 | 1 | 0.000722 | 0.000063 | 0.002145 | 0.000108 | 0.027730 | 0.001024 | Silicate | |
| M526-52 | A01 | 0.28 | 0.28 | 1 | 0.000617 | 0.000043 | 0.002023 | 0.000077 | 0.016295 | 0.007291 | Silicate | |
| M526-53 | A01 | 0.21 | 0.17 | 3 | 0.000345 | 0.000029 | 0.001617 | 0.000066 | 0.000427 | 0.008353 | Silicate | |
| M526-54 | A01 | 0.32 | 0.32 | 1 | 0.000544 | 0.000036 | 0.001406 | 0.000069 | 0.020324 | 0.000515 | Silicate | |
| M526-55 | A01 | 0.38 | 0.38 | 1 | 0.001360 | 0.000045 | 0.001915 | 0.000054 | 0.022564 | 0.005055 | Oxide | |
| M526-56 | A09 | 0.26 | 0.26 | 1 | 0.000633 | 0.000048 | 0.001906 | 0.000085 | 0.019643 | 0.013104 | Silicate | |
| M526-57 | A09 | 0.22 | 0.19 | 1 | 0.000680 | 0.000050 | 0.002093 | 0.000087 | 0.017602 | 0.007834 | Silicate | |
| M526-58 | A09 | 0.28 | 0.28 | 1 | 0.000584 | 0.000037 | 0.001856 | 0.000068 | 0.018401 | 0.007402 | Silicate | EDX |
| M526-59 | A09 | 0.28 | 0.28 | 1 | 0.000605 | 0.000040 | 0.001439 | 0.000073 | 0.023200 | 0.001633 | Silicate | |
| M526-60 | A09 | 0.22 | 0.18 | 1 | 0.000548 | 0.000050 | 0.001478 | 0.000095 | 0.022655 | 0.005679 | Oxide | |
| M526-61 | A09 | 0.31 | 0.31 | 1 | 0.000622 | 0.000036 | 0.001252 | 0.000061 | 0.017684 | 0.020077 | Silicate | |
| M526-62 | A09 | 0.25 | 0.25 | 1 | 0.000617 | 0.000057 | 0.001463 | 0.000074 | 0.012551 | 0.004245 | Silicate | |
| M526-63 | A09 | 0.29 | 0.29 | 3 | 0.000349 | 0.000030 | 0.001643 | 0.000067 | 0.022124 | 0.005415 | Silicate | |
| M526-64 | B02 | 0.19 | 0.14 | 1 | 0.000741 | 0.000075 | 0.001583 | 0.000123 | 0.020510 | 0.000452 | Silicate | EDX |
| M526-65 | B02 | 0.23 | 0.20 | 1 | 0.000778 | 0.000076 | 0.001919 | 0.000122 | 0.018832 | 0.000121 | Silicate | EDX |
| M526-66 | B02 | 0.28 | 0.28 | 2 | 0.000846 | 0.000051 | 0.000964 | 0.000098 | 0.030172 | 0.010553 | Silicate | EDX |
| M526-67 | B02 | 0.21 | 0.17 | 1 | 0.000822 | 0.000070 | 0.002009 | 0.000110 | 0.020582 | 0.004461 | Silicate | EDX |
| M526-68 | B03 | 0.26 | 0.26 | 1 | 0.000816 | 0.000065 | 0.001957 | 0.000101 | 0.021805 | 0.012337 | Silicate | EDX |
| M526-69 | B03 | 0.69 | 0.69 | 2 | 0.000978 | 0.000049 | 0.001083 | 0.000047 | 0.018598 | 0.021174 | Composite | EDX |
| M526-70 | B03 | 0.47 | 0.47 | 4 | 0.000515 | 0.000037 | 0.003734 | 0.000098 | 0.023834 | 0.005400 | Silicate | EDX |
| M526-71 | B03 | 0.22 | 0.18 | 4 | 0.000365 | 0.000053 | 0.002738 | 0.000143 | 0.017624 | 0.020194 | Silicate | EDX |
| M526-72 | B03 | 0.28 | 0.28 | 1 | 0.000883 | 0.000080 | 0.001036 | 0.000142 | 0.023743 | 0.002348 | Silicate | EDX |
| M526-73 | B04 | 0.26 | 0.26 | 1 | 0.000697 | 0.000064 | 0.001144 | 0.000091 | 0.030171 | 0.002260 | Silicate | EDX |
| M526-74 | B04 | 0.18 | 0.13 | 1 | 0.000781 | 0.000075 | 0.002120 | 0.000125 | 0.014674 | 0.007814 | Silicate | EDX |
| M526-75 | B04 | 0.31 | 0.31 | 1 | 0.001000 | 0.000061 | 0.001801 | 0.000087 | 0.030682 | 0.006387 | Silicate | EDX |
| M526-76 | B06 | 0.37 | 0.37 | 1 | 0.000866 | 0.000048 | 0.001876 | 0.000074 | 0.034679 | 0.007793 | Oxide | EDX |
| M526-77 | B06 | 0.24 | 0.21 | 4 | 0.000348 | 0.000049 | 0.002642 | 0.000128 | 0.026805 | 0.010162 | Silicate | EDX |
| M526-78 | B06 | 0.19 | 0.15 | 1 | 0.000557 | 0.000087 | 0.001126 | 0.000166 | 0.028079 | 0.027449 | Silicate | |
| M526-79 | B06 | 0.26 | 0.26 | 1 | 0.000685 | 0.000050 | 0.001462 | 0.000075 | 0.022794 | 0.013136 | Silicate | |
| M526-80 | C01 | 0.40 | 0.40 | 1 | 0.001162 | 0.000061 | 0.001942 | 0.000081 | 0.020819 | 0.005613 | Silicate | EDX |
| M526-81 | C03 | 0.19 | 0.15 | 1 | 0.000710 | 0.000069 | 0.001819 | 0.000118 | 0.017192 | 0.001541 | Silicate | EDX |
| N8276-01 | A01 | 0.28 | 0.28 | 1 | 0.000784 | 0.000058 | 0.001927 | 0.000092 | 0.018707 | 0.008461 | Silicate | EDX |
| N8276-02 | A07 | 0.26 | 0.26 | 1 | 0.000441 | 0.000031 | 0.001557 | 0.000065 | 0.012920 | 0.004774 | Silicate | |
| N8276-03 | A07 | 0.17 | 0.12 | 4 | 0.000424 | 0.000053 | 0.002677 | 0.000131 | 0.017974 | 0.002091 | Silicate | EDX |
| N8276-04 | A08 | 0.34 | 0.34 | 1 | 0.001506 | 0.000056 | 0.001754 | 0.000064 | 0.023719 | 0.010934 | Silicate | EDX |
| N8276-05 | A08 | 0.22 | 0.19 | 1 | 0.000667 | 0.000056 | 0.001738 | 0.000096 | 0.016447 | 0.012139 | Silicate | EDX |
| N8276-06 | A08 | 0.24 | 0.21 | 1 | 0.001108 | 0.000098 | 0.001658 | 0.000131 | 0.023549 | 0.004598 | Silicate | |
| N8276-07 | A14 | 0.24 | 0.20 | 1 | 0.001295 | 0.000074 | 0.001941 | 0.000091 | 0.015193 | 0.008282 | Silicate | |
| N8276-08 | B03 | 0.23 | 0.20 | 1 | 0.001457 | 0.000198 | 0.002074 | 0.000238 | 0.013926 | 0.005712 | Silicate | |

**Table S1:** B - Characteristics of presolar C-anomalous grains identified in Semarkona, MET 00526 and NWA 8276.

| Grain | Region | Size (μm)* | Size corr | $^{12}C/^{13}C$ | $\delta^{13}C$ | $\delta^{13}C$ error | $Si^-/C^-$ | Inferred Phase | Comment |
|---|---|---|---|---|---|---|---|---|---|
| SEM-SiC-01 | F1 | 0.26 | 0.26 | 68.8 | 293 | 38 | 0.18 | SiC | |
| SEM-SiC-02 | F1 | 0.28 | 0.28 | 62.0 | 436 | 48 | 0.28 | SiC | |
| SEM-C-01 | F1 | 0.77 | 0.77 | 134.3 | -337 | 12 | 0.18 | organics | clump of three $^{13}C$-poor grains |
| SEM-C-02 | F1 | 0.33 | 0.33 | 118.1 | -246 | 37 | 0.23 | organics | |
| SEM-SiC-03 | F2A1 | 0.38 | 0.38 | 9.5 | 8385 | 188 | 1.34 | SiC | |
| SEM-C-03 | F2A1 | 0.24 | 0.21 | 73.9 | 204 | 39 | 0.09 | organics | |
| SEM-SiC-04 | F2A1 | 0.31 | 0.31 | 64.6 | 379 | 60 | 0.35 | SiC | |
| SEM-C-04 | F2A1 | 0.22 | 0.18 | 76.9 | 158 | 33 | 0.05 | organics | |
| SEM-C-05 | F2A1 | 0.35 | 0.35 | 77.5 | 149 | 24 | 0.07 | organics | |
| SEM-C-06 | F2A1 | 0.40 | 0.40 | 72.7 | 225 | 40 | 0.32 | organics | |
| SEM-C-07 | F2A1 | 0.26 | 0.26 | 133.5 | -333 | 77 | 0.66 | organics | |
| SEM-SiC-05 | F2A1 | 0.30 | 0.30 | 53.3 | 669 | 41 | 0.92 | SiC | TEM, F2-30a in Singerling et al. (2022) |
| SEM-C-08 | F2A1 | 0.22 | 0.18 | 112.8 | -211 | 40 | 0.02 | organics | |
| SEM-C-09 | F2A1 | 0.17 | 0.12 | 128.1 | -305 | 61 | 0.07 | organics | |
| SEM-SiC-06 | F2A1 | 0.33 | 0.33 | 42.3 | 1105 | 77 | 0.47 | SiC | |
| SEM-C-10 | F2A1 | 0.31 | 0.31 | 109.3 | -186 | 34 | 0.01 | organics | |
| SEM-SiC-07 | F2A1 | 0.38 | 0.38 | 51.2 | 739 | 48 | 0.98 | SiC | |
| SEM-SiC-08 | F2A1 | 0.28 | 0.28 | 29.7 | 1998 | 148 | 0.92 | SiC | |
| SEM-C-11 | F2A1 | 0.32 | 0.32 | 76.0 | 171 | 36 | 0.19 | organics | |
| SEM-SiC-09 | F2A1 | 0.20 | 0.16 | 65.7 | 354 | 70 | 0.29 | SiC | |
| SEM-C-12 | F2A1 | 0.25 | 0.25 | 115.0 | -226 | 47 | 0.11 | organics | |
| SEM-SiC-10 | F2A1 | 0.28 | 0.28 | 66.4 | 341 | 68 | 0.38 | SiC | |
| SEM-SiC-11 | F2A2 | 0.17 | 0.12 | 61.0 | 459 | 102 | 0.55 | SiC | |
| SEM-SiC-12 | F2A2 | 0.38 | 0.38 | 59.5 | 496 | 28 | 0.81 | SiC | |
| SEM-SiC-13 | F2A2 | 0.30 | 0.30 | 49.9 | 785 | 91 | 0.64 | SiC | |
| SEM-C-13 | F2A2 | 0.30 | 0.30 | 79.3 | 122 | 27 | 0.08 | organics | |
| SEM-SiC-14 | F2A2 | 0.30 | 0.30 | 45.1 | 975 | 92 | 0.41 | SiC | |
| SEM-SiC-15 | F2A2 | 0.30 | 0.30 | 45.7 | 946 | 113 | 1.25 | SiC | |
| SEM-SiC-16 | F2A2 | 0.27 | 0.27 | 63.0 | 413 | 73 | 0.50 | SiC | |
| SEM-SiC-17 | F2A2 | 0.25 | 0.25 | 63.4 | 404 | 51 | 0.51 | SiC | |
| SEM-C-14 | F2A2 | 0.42 | 0.42 | 77.6 | 146 | 21 | 0.05 | organics | |
| SEM-C-15 | F2A2 | 0.40 | 0.40 | 109.0 | -184 | 21 | 0.17 | organics | |
| SEM-SiC-18 | F2A2 | 0.33 | 0.33 | 61.2 | 455 | 44 | 0.60 | SiC | |
| SEM-SiC-19 | F2A2 | 0.29 | 0.29 | 17.9 | 3963 | 211 | 1.01 | SiC | |
| SEM-SiC-20 | F2A2 | 0.24 | 0.20 | 44.4 | 1003 | 132 | 0.76 | SiC | |
| SEM-SiC-21 | F2A2 | 0.26 | 0.26 | 38.4 | 1316 | 136 | 0.84 | SiC | |
| M526-C-01 | AX | 0.46 | 0.46 | 73.0 | 219 | 28 | 0.22 | organics | out of focus, excluded from abundance calculation |
| M526-SiC-01 | AX | 0.45 | 0.45 | 62.0 | 436 | 77 | 1.79 | SiC | out of focus, excluded from abundance calculation |
| M526-SiC-02 | AX | 0.35 | 0.35 | 60.8 | 464 | 83 | 1.86 | SiC | out of focus, excluded from abundance calculation |
| M526-SiC-03 | AX | 0.60 | 0.60 | 62.2 | 432 | 25 | 1.25 | SiC | out of focus, excluded from abundance calculation |
| M526-SiC-04 | AX | 0.48 | 0.48 | 36.1 | 1465 | 40 | 1.11 | SiC | out of focus, excluded from abundance calculation |
| M526-SiC-05 | AX | 0.33 | 0.33 | 50.5 | 763 | 101 | 1.40 | SiC | out of focus, excluded from abundance calculation |
| M526-SiC-06 | AX | 0.23 | 0.20 | 47.2 | 887 | 182 | 2.12 | SiC | out of focus, excluded from abundance calculation |
| M526-SiC-07 | AX | 0.53 | 0.53 | 44.4 | 1006 | 60 | 1.61 | SiC | out of focus, excluded from abundance calculation |
| M526-SiC-08 | AX | 0.37 | 0.37 | 136.7 | -349 | 57 | 1.18 | SiC | out of focus, excluded from abundance calculation |
| M526-SiC-09 | A02 | 0.29 | 0.29 | 18.1 | 3917 | 293 | 2.68 | SiC | |
| M526-SiC-10 | A02 | 0.37 | 0.37 | 50.6 | 759 | 38 | 1.06 | SiC | |
| M526-C-02 | A02 | 0.29 | 0.29 | 76.3 | 166 | 35 | 0.05 | organics | |
| M526-SiC-11 | A02 | 0.28 | 0.28 | 13.8 | 5433 | 356 | 2.58 | SiC | |
| M526-SiC-12 | A02 | 0.25 | 0.22 | 20.7 | 3305 | 319 | 1.85 | SiC | |
| M526-SiC-13 | A05 | 0.22 | 0.18 | 47.2 | 884 | 198 | 1.19 | SiC | |
| M526-SiC-14 | A07 | 0.29 | 0.29 | 62.1 | 433 | 58 | 0.91 | SiC | |
| M526-C-03 | A07 | 0.23 | 0.20 | 211.7 | -580 | 93 | 0.47 | ambiguous | |
| M526-SiC-15 | A07 | 0.21 | 0.17 | 42.2 | 1111 | 177 | 1.16 | SiC | |
| M526-SiC-16 | A07 | 0.29 | 0.29 | 32.0 | 1785 | 160 | 1.37 | SiC | |
| M526-SiC-17 | A07 | 0.22 | 0.18 | 33.9 | 1624 | 244 | 1.27 | SiC | |
| M526-SiC-18 | A07 | 0.27 | 0.27 | 56.1 | 586 | 88 | 1.03 | SiC | |
| M526-SiC-19 | A07 | 0.34 | 0.34 | 44.3 | 1007 | 50 | 0.89 | SiC | |
| M526-SiC-20 | A07 | 0.26 | 0.26 | 50.7 | 756 | 84 | 0.87 | SiC | |
| M526-C-04 | A07 | 0.21 | 0.17 | 70.9 | 255 | 62 | 0.13 | organics | |
| M526-SiC-21 | A07 | 0.19 | 0.14 | 3.2 | 26860 | 537 | 1.19 | SiC | |
| M526-SiC-22 | A08 | 0.28 | 0.28 | 56.9 | 565 | 64 | 1.12 | SiC | |
| M526-C-05 | A11 | 0.30 | 0.30 | 56.4 | 579 | 61 | 0.09 | ambiguous | |
| M526-SiC-23 | A11 | 0.28 | 0.28 | 18.6 | 3791 | 245 | 0.51 | SiC | |
| M526-SiC-24 | A12 | 0.24 | 0.21 | 43.0 | 1071 | 162 | 0.22 | SiC | |
| M526-SiC-25 | A12 | 0.33 | 0.33 | 48.5 | 835 | 67 | 0.13 | SiC | |
| M526-SiC-26 | A13 | 0.28 | 0.28 | 7.8 | 10463 | 599 | 1.22 | SiC | |
| M526-SiC-27 | A14 | 0.21 | 0.17 | 53.5 | 662 | 129 | 0.27 | SiC | |
| M526-SiC-28 | A01 | 0.40 | 0.40 | 11.7 | 6606 | 205 | 0.37 | SiC | |
| M526-SiC-29 | A09 | 0.20 | 0.16 | 42.1 | 1115 | 154 | 0.23 | SiC | |
| M526-SiC-30 | A09 | 0.21 | 0.17 | 48.9 | 821 | 86 | 0.07 | SiC | |
| M526-SiC-31 | A09 | 0.30 | 0.30 | 61.4 | 449 | 74 | 0.16 | SiC | |
| M526-SiC-32 | A09 | 0.28 | 0.28 | 167.0 | -467 | 56 | 0.13 | SiC | |
| M526-SiC-33 | B02 | 0.24 | 0.21 | 5.1 | 16408 | 525 | 0.43 | SiC | |
| M526-SiC-34 | B04 | 0.24 | 0.20 | 10.0 | 7938 | 901 | 1.57 | SiC | |
| M526-SiC-35 | B04 | 0.18 | 0.13 | 22.3 | 2982 | 570 | 0.96 | SiC | |
| M526-SiC-36 | B04 | 0.15 | 0.08 | 50.5 | 761 | 203 | 0.22 | SiC | |
| M526-SiC-37 | B04 | 0.21 | 0.17 | 12.2 | 6293 | 303 | 0.43 | SiC | |
| M526-SiC-38 | B04 | 0.23 | 0.20 | 138.2 | -356 | 91 | 0.55 | SiC | |
| M526-C-06 | B04 | 0.23 | 0.20 | 143.2 | -378 | 82 | 0.07 | ambiguous | |
| M526-SiC-39 | B06 | 0.25 | 0.25 | 20.6 | 3313 | 401 | 0.99 | SiC | |
| M526-SiC-40 | B06 | 0.25 | 0.22 | 21.1 | 3215 | 295 | 0.67 | SiC | |
| M526-SiC-41 | B06 | 0.24 | 0.20 | 59.1 | 507 | 128 | 0.24 | SiC | |
| M526-SiC-42 | C01 | 0.21 | 0.17 | 36.2 | 1457 | 311 | 0.42 | SiC | |
| M526-SiC-43 | C01 | 0.17 | 0.12 | 18.1 | 3909 | 735 | 1.02 | SiC | |
| M526-SiC-44 | C02 | 0.22 | 0.18 | 181.7 | -510 | 120 | 0.15 | SiC | |
| M526-SiC-45 | C02 | 0.23 | 0.20 | 42.9 | 1075 | 142 | 0.35 | SiC | |
| M526-SiC-46 | C02 | 0.29 | 0.29 | 16.5 | 4401 | 160 | 0.16 | SiC | |
| M526-SiC-47 | C03 | 0.24 | 0.20 | 9.1 | 8769 | 681 | 0.89 | SiC | |
| M526-SiC-48 | C03 | 0.20 | 0.16 | 12.9 | 5895 | 923 | 1.81 | SiC | |
| M526-SiC-49 | C03 | 0.20 | 0.16 | 7.9 | 10230 | 1356 | 2.06 | SiC | |
| M526-C-07 | cluster (Fig. 1c) | | | | | | | | not counted as grain in the paper |
| N8276-SiC-01 | A04 | 0.25 | 0.25 | 58.3 | 527 | 58 | 0.04 | SiC | |
| N8276-SiC-02 | A05 | 0.27 | 0.27 | 6.1 | 13538 | 675 | 0.09 | SiC | |
| N8276-SiC-03 | A08 | 0.21 | 0.17 | 39.4 | 1257 | 279 | 0.24 | SiC | |
| N8276-SiC-04 | A11 | 0.26 | 0.26 | 45.7 | 946 | 114 | 0.22 | SiC | |
| N8276-C-01 | A11 | 0.24 | 0.20 | 11.5 | 6752 | 1078 | 0.80 | graphite | |
| N8276-SiC-05 | A13 | 0.25 | 0.25 | 7.7 | 10604 | 869 | 0.92 | SiC | |
| N8276-SiC-06 | A13 | 0.29 | 0.29 | 5.7 | 14632 | 426 | 0.05 | SiC | |
| N8276-SiC-07 | A01 | 0.25 | 0.21 | 41.1 | 1166 | 222 | 0.24 | SiC | |
| N8276-SiC-08 | B04 | 0.25 | 0.22 | 52.3 | 702 | 99 | 0.16 | SiC | |
| N8276-SiC-09 | B04 | 0.31 | 0.31 | 35.5 | 1505 | 83 | 0.12 | SiC | |

**Table S1 A+B Notes:**

Please find an .xlsx version of this table in the Supplementary Material.

*ROI-diameter. ROI sizes were defined as full width at half maximum. Displayed data was not corrected for beam size

EDX: Phase determined with help of EDX data. Phase determination may be unreliable.

$\delta^{13}C = [(^{13}C/^{12}C)_{grain}/(^{13}C/^{12}C)_{std} - 1] \times 1000$; $(^{13}C/^{12}C)_{std} = 0.011237$ (VPDB)

$\delta^{x}O = [(^{x}O/^{16}O)_{grain}/(^{x}O/^{16}O)_{std} - 1] \times 1000$; $x = 17, 18$; $(^{17}O/^{16}O)_{std} = 0.0003829$, $(^{18}O/^{16}O)_{std} = 0.0020052$ (VSMOW)

## NanoSIMS image simulations

The NanoSIMS ion probe is the instrument of choice for in situ searches of presolar grains, because most presolar silicates would be destroyed by other methods (e.g., during acid-dissolution) and because most grains are relatively small (<500 nm). The NanoSIMS can achieve spatial resolutions for isotopic mappings down to 50 nm (Hoppe et al. 2015) but due to tradeoffs between signal strength (and, hence, analysis time) and beam size, the typical beam sizes used to search for presolar silicate is ~100 nm. Unfortunately, even with high spatial resolution, the roughly Gaussian shaped primary ion beam of the NanoSIMS will always result in some signal arising from the surrounding material, thereby diluting the anomalous compositions of presolar grains with isotopically normal signal from their host matrix (e.g., Nguyen et al 2007, 2010). In situ NanoSIMS studies, therefore, measure narrower ranges of isotopic compositions than those seen in presolar oxides from acid residues (Zinner 2014). Furthermore, presolar grain sizes need to be determined accurately to estimate presolar grain abundances, but the apparent grain sizes seen in the ion images will generally be different from their real sizes. Small grains (i.e., <200 µm) are the most severely affected by isotope dilution and some of them might be missed entirely. In principle, it is possible to correct NanoSIMS imaging data for the biases introduced by isotopic dilution, for example by means of image simulations (Qin et al. 2011; Hoppe et al. 2021); however, no universally accepted procedure exists, making interlaboratory comparisons problematic.

To explore the impacts of isotopic dilution and counting statistics on the detectability, inferred sizes, and measured isotopic compositions of presolar grains, we performed three types of NanoSIMS image simulations: (i) idealized images with artificially introduced presolar grains of known size and isotopic composition (e.g., Nguyen et al. 2007), (ii) more realistic images to reproduce specific presolar grain candidates from the present study, and (iii) simulations based on real images but without anomalous grains in order to test false positive rates. All simulations were 256×256 pixel, 10×10 $\mu m^2$ images to match to mapping conditions used here. A brief summary of the modelling results and their implications can be found in the main text (Section 4.5.).

The idealized simulations are constructed similarly to those reported by Nguyen et al. (2007) and Hoppe et al. (2018). Namely, presolar grains with a range of isotopic enrichments or depletions and sizes were randomly placed within homogenous $^{16}O$ and $^{17}O$ images and smoothed by an assumed Gaussian-shaped beam of 100-nm diameter. Very long counting times were assumed so that counting statistical errors are negligible. These simulations were then analyzed in the same manner as the real meteoritic images, with the simulated presolar grains defined based on sigma images as described in Section 2. For each simulated grain, we derived the measured grain size and $^{17}O/^{16}O$ ratio. Shown as grey symbols in Fig. S1a are the true isotopic ratios of the simulated grains plotted against their measured ratios. Grey solid lines connect grains of the same size, while the grey dashed lines indicate constant degrees of dilution (i.e., a dilution value of 40% means that 40% of the measured signal comes from surrounding pixels). As expected, the degree of isotopic dilution is seen to decrease with increasing grain size. We also note that, for a given grain size, there is not a simple relationship between the degree of dilution and the magnitude of the true anomaly, as indicated by the non-parallelism of the solid and dashed grey curves.

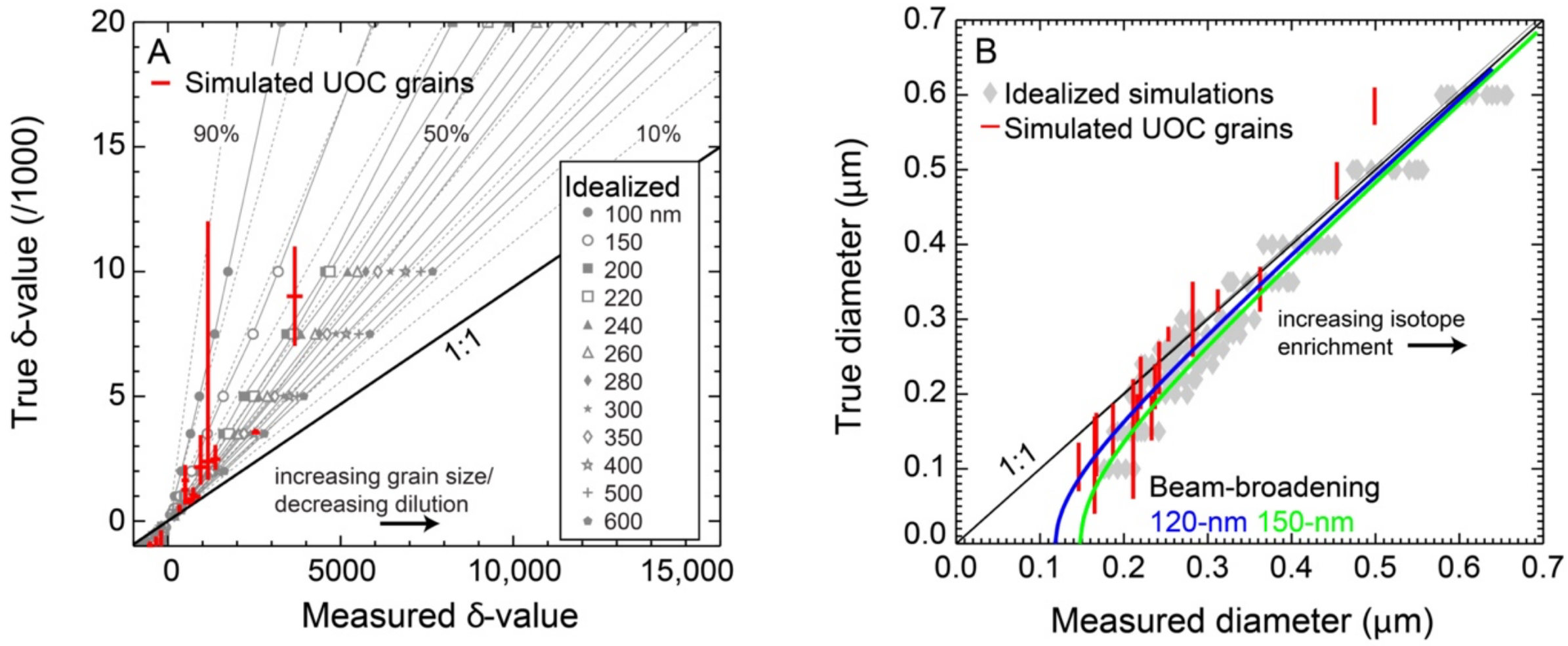

**Fig. S1:** Results of NanoSIMS image simulations. Grey symbols represent idealized simulations in which simulated presolar grains with a range of isotopic enrichments or depletions and sizes were randomly placed within homogenous $^{16}O$ and $^{17}O$ images. Red bars indicate more realistic simulations aimed at reproducing specific presolar grains (see text). A) The true input isotopic ratios of the simulated grains are plotted against the ratios derived from analysis of the simulated images ("measured"). Grey solid lines connect grains of the same size. Grey dashed lines indicate constant degree of isotopic dilution. The degree of isotopic dilution is seen to decrease with increasing grain size. Vertical red bars indicate the range of input isotopic ratios which can provide a good match to real presolar grains for which horizontal bars represent the measured value and one-sigma errors (Table S2). B) The true input presolar grain sizes are plotted against the apparent (measured) grain diameters. The blue curve indicates the effect of adding a 120-nm broadening due to ratio image smoothing in quadrature to the true grain sizes, whereas the green curve additionally includes the blurring effect of the 100-nm Gaussian primary beam (see text). Red vertical bars indicate range of input sizes that match observed data.

The true input presolar grain sizes are plotted against the apparent (measured) grain diameters as grey symbols in Fig. S1b. For grains larger than ~250 nm, the data scatter around the 1:1 line; smaller grains generally appear slightly larger than their true sizes. In fact, the minimum apparent diameter for any of the simulated presolar grains is ~180 nm despite the simulations including grains of size 100 and 150 nm. We attribute this to the combined effect of the finite primary ion beam width (100 nm) and the boxcar smoothing applied to the images prior to the generation of isotopic ratio images. We used a 3×3 pixel smoothing window, corresponding to a blurring of ~120 nm. The blue curve in Fig. S1b indicates the effect of adding this size in quadrature to the true grain sizes, whereas the green curve additionally includes the blurring effect of the Gaussian primary beam. While these curves can clearly explain the average size of the modeled presolar grains, the large scatter around them indicates that a simple correction to measured grain sizes as performed in our previous studies is in fact unwarranted. Note also that for a given true grain size, the apparent diameter increases with increasing isotopic enrichment. This phenomenon can be understood based on our method of defining presolar grain ROIs in "sigma" images, where each pixel value is assigned the number of $\sigma$ its isotopic ratio is from normal. Since for Poisson statistics $\sigma$ is given by the square root of the number of counts, more isotopically enriched grains have more pixels above a given significance threshold than less anomalous grains of the same true size and thus appear larger.

In principle, one could invert simulation results like those shown in Fig. S1 to infer true isotopic ratios and grain sizes from measured ones, but this is made difficult by counting statistical errors in real images. Under typical measurement conditions, the errors in measured isotopic ratios are such that a wide range of models could be consistent with the observed data. To investigate this in more detail, we performed more realistic image simulations of several presolar grain candidates from our study (Table S2). For each grain simulation, we used the observed $^{16}O$ image as a template from which to draw $^{16}O$, $^{17}O$ and $^{18}O$ pixel counts from Poisson distributions (Fig. S2). Images were assumed to be isotopically normal except for artificially-placed presolar grains of a given isotopic composition and size, and convolved with a 100-nm Gaussian primary beam. We varied the presolar grain parameters (composition and size) over wide ranges. For each composition/size pair we generated 5 to 10 simulations, defined presolar grain ROIs in each (if anomalies were detectable), and calculated the average measured isotopic ratio and size. In this way we could estimate the range of true grain sizes and compositions that result in images that match the real data.

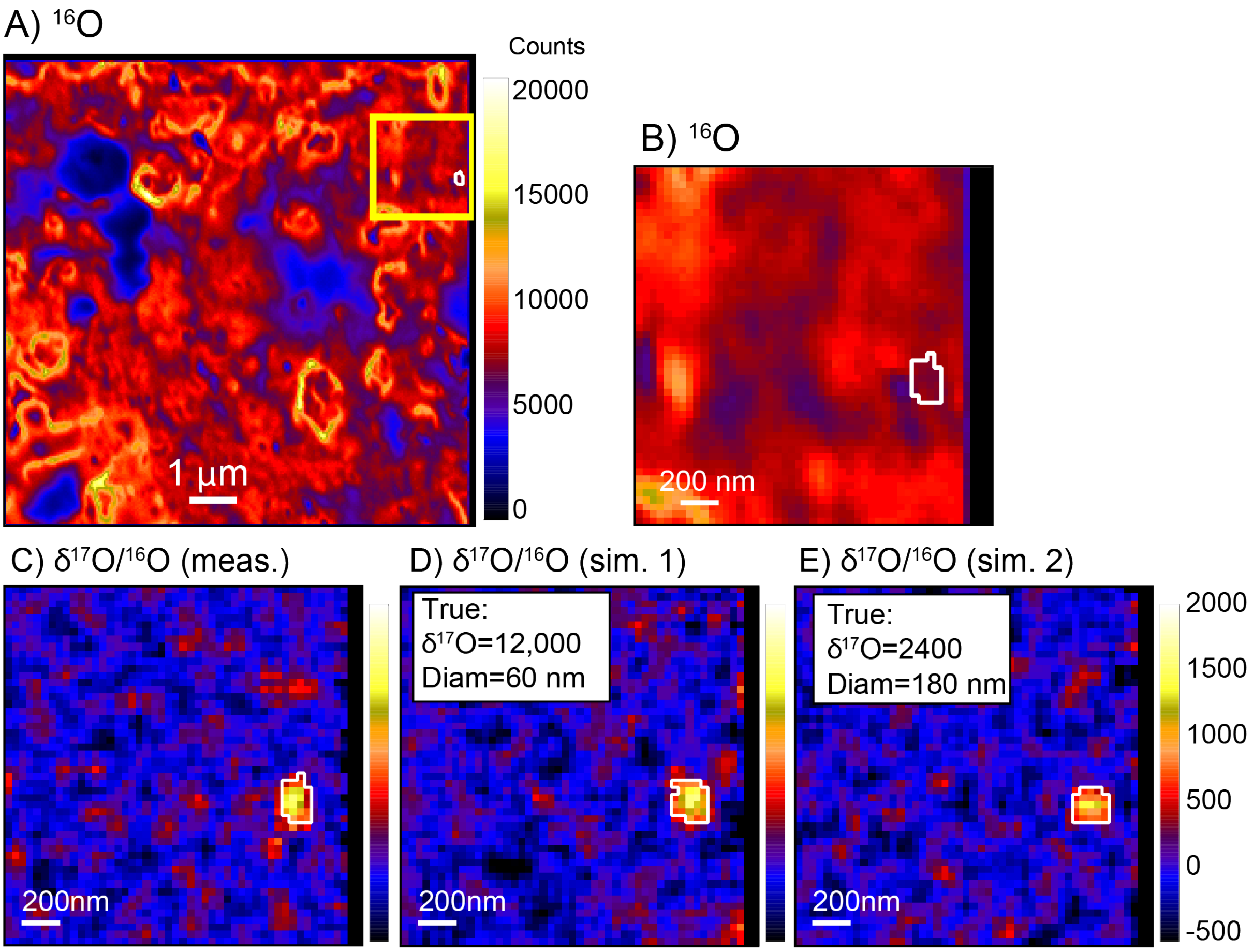

**Fig. S2:** NanoSIMS image simulations to match the $^{17}$O-rich presolar grain M526-67. A, B) Original $^{16}$O ion image for grain. C) Measured and D, E) simulated $\delta^{17}O/^{16}O$ images of the same grain with indicated true input characteristics. Both simulations match the observed anomaly and grain size very well (see Table S2), despite very different simulation parameters.

Figure S2 shows example simulations for $^{17}$O-rich presolar grain (M526-67). The top panels show the original NanoSIMS $^{16}O^-$ image in which this grain was identified, while the bottom ones show the measured and two simulated $^{17}O/^{16}O$ images for a zoomed-in area around the grain. Both simulations matched the observed anomaly very well in its apparent size and composition. However, as indicated, these represent two very different sets of input parameters, showing that grains with very different sizes and isotopic compositions can appear the same in NanoSIMS images, under our imaging conditions. In fact, in cases like this of small, isotopically enriched grains, there is generally a tradeoff in which smaller, more extremely anomalous grains will appear the same as larger, less-extreme grains, though this example is a somewhat extreme case.

**Table S2:** Presolar grain simulations based on real grains identified in this study.

| | Measured | | | | Simulated | | | |
|---|---|---|---|---|---|---|---|---|
| Grain | diameter (μm) | δ-value (‰) | | | min. diam | max. diam. | min. δ-value (‰) | max. δ-value (‰) |
| M526-36 | 0.45 | 851 | ± | 71 | 0.46 | 0.51 | 900 | 1000 |
| M526-39 | 0.24 | 1364 | ± | 146 | 0.18 | 0.24 | 2000 | 3000 |
| M526-40 | 0.22 | 480 | ± | 51 | 0.18 | 0.25 | 650 | 850 |
| M526-41 | 0.25 | 617 | ± | 113 | 0.27 | 0.29 | 650 | 900 |
| M526-46 | 0.23 | 3681 | ± | 231 | 0.14 | 0.20 | 7000 | 11000 |
| M526-48 | 0.28 | 718 | ± | 114 | 0.25 | 0.35 | 700 | 1300 |
| M526-55 | 0.36 | 2537 | ± | 118 | 0.31 | 0.37 | 3450 | 3650 |
| M526-61 | 0.31 | 605 | ± | 89 | 0.31 | 0.34 | 600 | 900 |
| M526-67 | 0.21 | 1146 | ± | 183 | 0.06 | 0.22 | 1600 | 12000 |
| M526-70 | 0.50 | 823 | ± | 45 | 0.56 | 0.61 | 900 | 950 |
| M526-77 | 0.24 | 317 | ± | 64 | 0.20 | 0.27 | 300 | 600 |
| MET-B03_8 | 0.17 | 483 | ± | 106 | 0.04 | 0.17 | 600 | 2200 |
| MET-B04_6 | 0.15 | 481 | ± | 105 | 0.07 | 0.14 | 900 | 1500 |
| M526-70 | 0.17 | -350 | ± | 58 | 0.10 | 0.15 | -1000 | -600 |
| M526-78 | 0.19 | -439 | ± | 83 | 0.13 | 0.20 | -1000 | -700 |
| MET-A2_11 | 0.17 | -210 | ± | 39 | 0.09 | 0.18 | -1000 | -350 |
| NWA_A13_33 | 0.22 | -523 | ± | 87 | 0.16 | 0.20 | -1000 | -800 |

The results of these more realistic image simulations for 17 presolar grains are overlaid on the idealistic dilution models in Fig. S1 (red bars). The vertical errors represent the range of input parameters that provide a good match to the real data for each grain. Comparison with the models shows that, in general, the effects of isotopic dilution can be highly non-unique and one should proceed with caution in correcting derived compositions and sizes of presolar grains based solely on NanoSIMS images themselves, especially from idealized models. If additional information can be obtained (e.g., exact grain sizes from electron microscopy or higher-resolution NanoSIMS imaging), one may be more confident in correcting for isotope dilution effects. The three modeled grains with isotopic depletions are all consistent with a true δ-value for the presolar grain of -1000 ($^{17}O/^{16}O$ or $^{18}O/^{16}O$ ratio of zero), indicating that many previously reported grains with isotopic depletions may in fact be substantially more anomalous than recognized. For example, many (strongly $^{18}O$-depleted) Group 2 presolar O-anomalous grains found in situ have likely been misidentified as Group 1 grains. With regards to this point, we found a few isotopically depleted presolar grain candidates that could not be well simulated even with a true isotopic ratio of zero. All of these were near the significance threshold of 5σ so we concluded that their apparent anomalies were statistical flukes and excluded them from the data set.

In terms of grain size, Fig S1b shows that, for measured grain sizes larger than ≈250 nm, the realistic simulations are scattered around the 1:1 line indicating no need for any kind of beam broadening correction. In contrast, smaller grains appear systematically larger than their true sizes and scatter around the 120-nm beam broadening curve. We therefore only corrected grains <250 nm for such broadening as discussed above. It must be recognized that the uncertainty in such a correction is large due to the wide range of simulations that match each grain.

Finally, to investigate the possibility of false candidates arising from counting-statistical variations, we performed additional simulations in which we used real $^{16}O$ or $^{12}C$ images from the present data set as templates and generated simulations under the assumption that images contained no isotopically anomalous material (i.e., these were similar to the realistic simulations of specific grains described above except that no presolar grains were added). We analyzed 70 simulated O-isotopic images and 100 simulated C isotopic images. Analyses of these images via the same methodology used to analyze the real meteoritic data revealed a number of small regions with spurious isotopic anomalies in both sets of simulations. Based on the sizes and significance of these, we chose the significance thresholds described in Section 2.

**References**

Hoppe P., Leitner J. and Kodolányi J. (2015) New constraints on the abundances of silicate and oxide stardust from supernovae in the Acfer 094 meteorite. *Astrophys. J.* **808**, 9–15.

Hoppe P., Leitner J. and Kodolányi J. (2018) New insights into the galactic chemical evolution of Magnesium and Silicon isotopes from studies of silicate stardust. *Astrophys. J.* **869**, 47–60.

Hoppe P., Leitner J., Kodolányi J. and Vollmer C. (2021) Isotope systematics of presolar silicate grains: New insights from Magnesium and Silicon. *Astrophys. J.* **913**, 10–27.

Nguyen A. N., Stadermann F. J., Zinner E., Stroud R. M., Alexander C. M. O'D. and Nittler L. R. (2007) Characterization of presolar silicate and oxide grains in primitive carbonaceous chondrites. *Astrophys. J.* **656**, 1223–1240.

Nguyen A. N., Nittler L. R., Stadermann F. J., Stroud R. M. and Alexander C. M. O'D. (2010) Coordinated analyses of presolar grains in Allan Hills 77307 and Queen Elizabeth Range 99177 meteorites. *Astrophys. J.* **719**, 166–189.

Singerling S. A., Nittler L. R., Barosch J., Dobrică E., Brearley A. J. and Stroud R. M. (2022) TEM analyses of in situ presolar grains from unquilibrated ordinary chondrite LL3.0 Semarkona. *Geochim. Cosmochim. Acta* **328**, 130–152.

Qin L., Nittler L. R., Alexander C. M. O'D., Wang J., Stadermann F. J. and Carlson R. W. (2011) Extreme $^{54}Cr$-rich nano-oxides in the CI chondrite Orgueil – Implications for a late supernova injection into the solar system. *Geochim. Cosmochim. Acta* **75**, 629–644.

Zinner E. (2014) Presolar grains. In: *Meteorites and Cosmochemical Processes* (ed. A. M. Davis), Vol. 1, Treatise on Geochemistry 2nd ed. (exec. eds. H. D. Holland & K. K. Turekian), 181–213.